\documentclass[aoas,MSNbibl,nameyear,rotating,dvips]{arximspdf}
\usepackage{multirow}
\usepackage{graphicx}

\doi{10.1214/13-AOAS631} %kopijuoti is PTS
\volume{7}
\issue{3}
\pubyear{2013}
\firstpage{1421}
\lastpage{1449}

\makeatletter

\def\bm{\bolds}
\newcommand{\eqref}[1]{(\ref{#1})}
\makeatother

\begin{document}
\begin{frontmatter}

\title{Spatio-temporal exceedance locations and confidence regions\thanksref{T1}}
\runtitle{Exceedance locations and confidence regions}

\begin{aug}
\author[a]{\fnms{Joshua P.} \snm{French}\corref{}\ead[label=e1]{joshua.french@ucdenver.edu}}
\and
\author[b]{\fnms{Stephan R.} \snm{Sain}\ead[label=e2]{ssain@ucar.edu}}

\thankstext{T1}{Supported under NSF Grant ATM-0534173.}
\runauthor{J.~P. French and S.~R. Sain}
\affiliation{University of Colorado Denver and National Center\break for
Atmospheric~Research}
\address[a]{Department of Mathematical\\
\quad and Statistical Sciences\\
University of Colorado Denver\\
Campus Box 170\\
PO Box 173364\\
Denver, Colorado 80217-3364\\
USA\\
\printead{e1}}
\address[b]{National Center for Atmospheric Research\\
1850 Table Mesa Dr \\
Boulder, Colorado 80305\\
USA\\
\printead{e2}}
\end{aug}

% HISTORY:
\received{\smonth{2} \syear{2012}}
\revised{\smonth{1} \syear{2013}}

% ABSTRACT
%
\begin{abstract}
An exceedance region is the set of locations in a spatial domain where
a process exceeds some threshold. Examples of exceedance regions include
areas where ozone concentrations exceed safety standards, there is
high risk for tornadoes or floods, or heavy-metal levels are dangerously
high. Identifying these regions in a spatial or spatio-temporal setting
is an important responsibility in environmental monitoring. Exceedance
regions are often estimated by finding the areas where predictions
from a statistical model exceed some threshold. Even when estimation
error is quantifiable at individual locations, the overall estimation
error of the estimated exceedance region is still unknown. A method
is presented for constructing a confidence region containing the true
exceedance region of a spatio-temporal process at a certain time.
The underlying latent process and any measurement error are assumed
to be Gaussian. Conventional techniques are used to model the spatio-temporal
data, and then conditional simulation is combined with hypothesis
testing to create the desired confidence region. A simulation study
is used to validate the approach for several levels of spatial and
temporal dependence. The methodology is used to identify regions of
Oregon having high precipitation levels and also used in comparing
climate models and assessing climate change using climate models from
the North American Regional Climate Change Assessment Program.
\end{abstract}

% KEYWORDS

\end{frontmatter}

%s1 #&#
\section{Introduction}

Identifying regions of extreme or unusual response is often a vital
concern when analyzing environmental data. Identification of these
regions can have important health, societal, and political impacts
since these areas may indicate unusual events, outbreak of disease,
toxic conditions, regions of extreme risk, etc. [\citet
{Patil2010DigitalGovernance}].
These regions of interest are often described as exceedance regions
or hotspots, and are defined as the set of locations where the response
process of interest exceeds some specified threshold. On the basis
that the broad definition of exceed is ``to go beyond the bounds
or limits of'' something [\citet{Dictionary2012}], an exceedance
region may refer to the area where the responses are above a threshold
or the area where the responses fall below a threshold (though the
two settings should be carefully distinguished).

The uncertainty associated with estimating exceedance regions has
previously received relatively little attention. Our goal in this
paper will be to introduce an approach for identifying exceedance
regions with a quantifiable amount of certainty, and then using this
methodology to confidently identify exceedance regions in two different
contexts. First, in Section \ref{secApplication-1}, this methodology
will be used to confidently identify the regions of Oregon having
extreme precipitation levels in October of 1998 using data from the
previous two years. Being able to predict seasonal climate can help
to minimize ``climate surprises,'' reduce impacts on society and
ecosystems, assess the chances of events like drought and wildfire,
and make important economic decisions in fields such as agriculture,
energy, insurance, and public health [\citet{NOAA2011Predicting}].
In Section \ref{secApplication-2}, we use the proposed methodology
to explore the temperature data of climate models from the North American
Regional Climate Change Assessment Program (NARCCAP). Discussion will
include assessment of future climate change for several regions of
North America based on these models and also highlight similarities
and differences between the projections from various climate models.

The methodology proposed in this article allows one to construct a
confidence region for the exceedance region of a spatio-temporal process
at a certain time. The resulting confidence region will contain the
entire exceedance region with known confidence. To our knowledge,
no methods have previously been available for constructing confidence
regions containing the entire exceedance region of a spatial/spatio-temporal
process and having the desired coverage properties. The methodology
combines optimal spatio-temporal prediction methods and a
hypothesis-testing-like
approach to construct the confidence regions. Due to the simultaneous
nature of the inference, the size and shape of the confidence regions
will be directly related to the domain of interest. \citet
{French2010Level-Curves}
introduced an approach for identifying the level curves of a spatial
process with known confidence. Our methodology generalizes that of
\citet{French2010Level-Curves} by handling exceedance regions for
spatio-temporal processes and also accounting for additional uncertainty
when estimating the mean structure of the spatio-temporal process.
The exceedance region of a spatial or spatio-temporal process is often
predicted as the region(s) where the predicted response exceeds the
desired threshold. This approach results in a biased predicted exceedance
region since spatial predictors tend to oversmooth the response surface
[\citet{Zhang2008Loss}]. \citet{Craigmile2006Loss} accounted
for the oversmoothing of traditional spatial predictors by using a
loss function that assigned more weight to extreme values in prediction;
the predicted exceedance regions resulting from this approach are
often larger than those of traditional methods. \citet{Patil2004Upper}
used the upper level set (ULS) scan statistic, a modification of the
popular spatial scan statistic [\citet{Kulldorff1995Spatial};
\citet{Kulldorff1997Spatial}] to identify regions or clusters
of cells with elevated responses when compared to neighboring cells.
The spatial scan statistic is often used in the setting where a region
is tessellated into cells and response data are available as counts.
\citet{Patil2004Upper} used maximum likelihood estimation to identify
regions with elevated counts through exhaustive search of the parameter
space. The ULS scan statistic reduced the computational complexity
of their task by reducing the search space, which allowed \citeauthor{Patil2004Upper}
to construct a type of confidence region for the upper level sets
(hotspots) by finding all zones where the ULS statistic was not statistically
significant. The Progressive Upper Level Set (PULSE) scan statistic
was recently proposed by \citet{PatilEtAl2010PULSE} as a refinement
of the ULS statistic for detecting geospatial hotspots. Neither of
the previous two approaches incorporates the typical notion of spatial
dependence used to model and predict spatial data as exposited by
\citet{Cressie1993Statistics}, \citet{Schabenberger2005Statistical},
and many others. \citet{Zhang2008Loss} predicted the exceedance
region by finding the region minimizing the posterior loss of an image-based
loss function using simulated annealing. The predicted exceedance
region found by the approach of \citet{Zhang2008Loss} is statistically
optimal, but is not a confidence set in the traditional statistical
sense. Recently, \citet{Sun2012False} developed a unified theoretical
and computational framework for false discovery control in multiple
testing when trying to identify locations where the mean response
of a process exceeds some threshold.

Other subject areas with connections to exceedance regions include
the estimation of spatial cumulative distribution functions (CDFs)
and the analysis of fMRI experiments to identify activated voxels.
The spatial CDF of a process over a domain is a random distribution
function that provides a statistical summary of a random field over
a given region [\citet{lahiri1999asymptotic}]. Spatial CDFs were
introduced by \citet{Majure1996Spatial} and \citet{lahiri1999asymptotic},
and \citet{Lahirietal1999SCDF} used subsampling approaches to predict
a spatial CDF and study the asymptotic properties of the associated
predictors. \citet{Zhu2002Asymptotic} developed an approach to
detect a change in the spatial CDF of a region over time using the
difference between two empirical CDFs and then quantified any change
using the weighted integrated squared distance between the empirical
CDFs. While spatial CDFs provide a way of identifying the proportion
of a region exceeding some threshold, they do not identify where these
exceedances occur. In the medical imaging field, there has long been
an interest in identifying the areas of an fMRI map depicting brain
activity in response to some stimuli. \citet{Marchini2004Comparing}
provide a helpful overview of the various approaches to detecting
active voxels. The most common approach for identifying active voxels
is to declare a voxel active if some associated test statistic is
above a single common threshold. This threshold is determined in a
way that allows for control of some error criterion, such as the standard
familywise error rate (FWER) or False Discovery Rate [FDR, \citet
{Benjamini1995Controlling}].
Often, the approach to determining this threshold utilizes the theory
of random fields [\citet{Adler1981Geometry}, \citet{Adler2007Random}] and
the expected Euler characteristic
[\citet{Taylor2007Maxima,Adler2008Some}]
to determine the probability that the maximum test statistic is greater
than some threshold. This approach can naturally apply in a geostatistical
context such as the present one, but assumptions regarding the stationarity
of the test statistics can be overly strict and often produces conservative
results [\citet{Marchini2004Comparing}].

%s1.1 #&#
\subsection{Outline}

The approach proposed in this paper fills a methodological gap in
the literature by allowing researchers to construct confidence regions
for exceedance regions using traditional geostatistical tools and
notions of spatial dependence while having the confidence level properties
typically desired. The underlying latent process and any measurement
error are assumed to be Gaussian.

This paper will proceed in the following manner. In Section \ref{secMethodology}
we will describe the proposed methodology in detail. Section \ref
{secSimulation-Studies}
presents the results of a simulation study used to validate our approach,
as well as several case studies exploring the shape of the confidence
region for various underlying mean structures. Sections~\ref{secApplication-1}
and~\ref{secApplication-2} describe applications of this methodology
in the climatological settings mentioned above. We conclude with a
brief discussion in Section \ref{secDiscussion}.

%s2 #&#
\section{Methodology}\label{secMethodology}

%s2.1 #&#
\subsection{Framework}\label{subFramework}

\global\long\def\bs{\mathbf{s}}
\global\long\def\st{(\mathbf{s},t)}
\global\long\def\t{\mathsf{T}}
Consider a spatio-temporal response process $Y(\cdot,\cdot)\equiv \{
Y(\mathbf{s},t),(\mathbf{s},t)\in D\times T\subset\mathbb{R}^{2}\times
\mathbb{R}\}$,
which has spatial index $\bs$ contained in a bounded two-dimensional
region of interest $D$, and time index $t$ which is contained in
a bounded set of possible times $T\subset\mathbb{R}$. We assume that
$Y(\cdot,\cdot)$ is the sum of a hidden spatio-temporal process $Z(\cdot,\cdot)$
and a measurement error process $\varepsilon(\cdot,\cdot),$ so that
\[
Y\st=Z\st+\varepsilon\st,\qquad \st\in D\times T.
\]
We will refer to $Y(\cdot,\cdot)$ as the observable process and
$Z(\cdot,\cdot)$ as the hidden or latent process. The hidden 
process\vadjust{\goodbreak}
$Z(\mathbf{s},t)$ is composed of a mean structure $\mu\st$ capturing
large-scale response behavior and a mean-zero spatio-temporal process
$W\st$ capturing small-scale behavior, such that $Z\st$ can be decomposed
into
\[
Z\st=\mu\st+W\st, \qquad\st\in D\times T.
\]
The mean structure $\mu\st$ is assumed to follow the linear model
$\mu\st=\mathbf{x}\st^{\t}\bm{\beta},$ where $\mathbf{x}\st$ is
a $k\times1$ vector of known space--time covariates for location $\bs$
at time $t$, $\bm{\beta}$ is a vector of space--time trend parameters,
and $W\st$ is assumed to be a Gaussian process having continuous
sample paths [cf. \citet{Paciorek2003Nonstationary}, page~57; \citet
{Abrahamsen1997RandomFields}, \citet{Adler2007Random}].
The covariance between two responses for the hidden process is denoted
by
\[
\operatorname{cov}\bigl\{Z(\mathbf{s}_{1},t_{1}),Z(
\mathbf{s}_{2},t_{2})\bigr\} =C(\mathbf{s}_{1},
\mathbf{s}_{2},t_{1},t_{2}).
\]
The error process $\varepsilon\st$ is assumed to be a Gaussian white-noise
process with mean 0 and variance $\sigma_{\varepsilon}^{2}$, and the
covariance between two responses of the observable process is given
by
\[
\operatorname{cov}\bigl\{Y(\mathbf{s}_{1},t_{1}),Y(
\mathbf{s}_{2},t_{2})\bigr\} =C(\mathbf{s}_{1},
\mathbf{s}_{2},t_{1},t_{2})+\sigma_{\varepsilon
}^{2}v_{\varepsilon}(
\mathbf{s}_{1},\mathbf{s}_{2},t_{1},t_{2}),
\]
where $v_{\varepsilon}$ is a known function of the spatial and temporal
indices (often returning 1 when $\mathbf{s}_{1}=\mathbf{s}_{2}$
and $t_{1}=t_{2}$ and zero otherwise).

Suppose we have observed $n$ responses $\mathbf{y}=[Y(\mathbf{s}_{1},t_{1}),\ldots,Y(\mathbf{s}_{n},t_{n})]^{\t}$
of a partial realization of $Y(\cdot,\cdot)$. Our goal is to use
the observed responses to construct a confidence region for the locations
where the hidden process $Z(\cdot,\cdot)$ exceeds some threshold
$u$ at some time $t_{p}$. We define the exceedance region above
a threshold $u$ for the hidden process $Z$ as $E_{u^{+}}^{Z}=\{\mathbf{s}\in D\dvtx Z(\mathbf{s},t_{p})\geq u\}$,
while the exceedance region below a threshold $u$ may be defined
as $E_{u^{-}}^{Z}=\{\mathbf{s}\in D\dvtx Z(\mathbf{s},t_{p})\leq u\}$.
Depending on the exceedance region of interest (above or below the
threshold), we would like to find a region $S_{u^{+}}$ that contains
$E_{u^{+}}^{Z}$ with probability $1-\alpha$, that is,
$P(E_{u^{+}}^{Z}\subseteq S_{u^{+}})=1-\alpha$
or a region $S_{u^{-}}$ that contains $E_{u^{-}}^{Z}$ with probability
$1-\alpha$, that is, $P(E_{u^{-}}^{Z}\subseteq S_{u^{-}})=1-\alpha$.

%s2.2 #&#
\subsection{Methodology}\label{subMethodology-for-Calpha}

For simplicity, we will only describe methodology for determining
the confidence region $S_{u^{+}}$ (the methodology for determining
$S_{u^{-}}$ being analagous). Our approach for constructing our confidence
region $S_{u^{+}}$ will combine geostatistical techniques along with
an approach similar to hypothesis testing. We will state a null and
alternative hypothesis, determine a test statistic and critical value,
then make our conclusion. However, an important difference is that
these tests are related to random variables instead of parameters,
violating the classical definition of a hypothesis test. In spite
of this difference, we will use the traditional terminology of hypothesis
testing since the same concepts apply. Additionally, the procedure
will call for a large number of statistical tests so the critical
value must be adjusted in order to control the confidence level of
our confidence region. We note that though this methodology is presented
in the context of spatio-temporal processes, the approach is equally
applicable to purely spatial data (with no time varying component)
simply by assuming we have only one time.

The confidence region $S_{u^{+}}$ can be constructed by testing for
each $\mathbf{s}\in D$, $H_{0}\dvtx Z(\mathbf{s},t_{p})=u$ versus
$H_{a}\dvtx Z(\mathbf{s},t_{p})<u$
on the basis of a test statistic $Z^{\prime}(\mathbf{s},t_{p})$. The region
$S_{u^{+}}$ is made up of all locations $\mathbf{s}$ where we fail
to conclude that $Z(\mathbf{s},t_{p})<u$. Naturally, the test statistic
$Z^{\prime}(\mathbf{s},t_{p})$ is based on a predictor of $Z(\mathbf{s},t_{p})$,
$\hat{Z}(\mathbf{s},t_{p})$, from the statistical model outlined
in Section \ref{subFramework}. When the trend parameter vector $\bm
{\beta}$
must be estimated, the unbiased linear predictor of $Z\st$ minimizing
the mean-squared prediction error is known as the universal kriging
predictor. Suppose we wish to make a prediction of $Z(\cdot,\cdot)$
at location $\mathbf{s}_{0}$ and time $t_{0}$. Let $\mathbf{x}_{0}$
denote the $k\times1$ vector of space--time covariates associated
with $Z(\mathbf{s}_{0},t_{0})$, $\mathbf{X}_{\mathbf{y}}$ denote
the $n\times k$ matrix of covariates for the observed data $\mathbf{y}=\{Y(\mathbf{s}_{1},t_{1}),\ldots,Y(\mathbf{s}_{n},t_{n})\}$,
$\bolds{\Sigma}_{\mathbf{y}}$ denote the $n\times n$ covariance
matrix for the observed data $\mathbf{y}$, and
\[
\mathbf{c}_{0\mathbf{y}}=\bigl[C(\mathbf{s}_{0},\mathbf{s}_{1},t_{0},t_{1}),\ldots,C(
\mathbf{s}_{0},\mathbf {s}_{n},t_{0},t_{n})
\bigr]^{\t}
\]
be the $n\times1$ vector of covariances between the response to be
predicted and the observed data responses. The optimal predictor for
$Z(\mathbf{s}_{0},t_{0})$ is
%
%e1 #&#
\begin{eqnarray}\label{equk-prediction}
\hat{Z}(\mathbf{s}_{0},t_{0}) & =&\bolds{
\Sigma}_{\mathbf{y}}^{-1}\bigl(\mathbf{c}_{0\mathbf{y}}-
\mathbf{X}_{\mathbf{y}}\bigl(\mathbf {X}_{\mathbf{y}}^{\t}\bolds{
\Sigma}_{\mathbf{y}}^{-1}\mathbf {X}_{\mathbf{y}}\bigr)^{-1}
\bigl(\mathbf{X}_{\mathbf{y}}^{\t}\bm{\Sigma}_{\mathbf
{y}}^{-1}
\mathbf{c}_{0\mathbf{y}}-\mathbf{x}_{0}\bigr)\bigr)\mathbf{y}
\nonumber
\\[-8pt]
\\[-8pt]
\nonumber
& =&\mathbf{x}_{0}^{\t}\hat{\bm{\beta}}_{\mathrm{gls}}+
\mathbf{c}_{0\mathbf
{y}}^{\t}\bolds{\Sigma}_{\mathbf{y}}^{-1}(
\mathbf{y}-\mathbf {X}_{\mathbf{y}}\hat{\bm{\beta}}_{\mathrm{gls}}),
\nonumber
\end{eqnarray}
where $\hat{\bm{\beta}}_{\mathrm{gls}}=(\mathbf{X}_{\mathbf{y}}^{\t
}\bolds{\Sigma}_{\mathbf{y}}^{-1}\mathbf{X}_{\mathbf{y}})^{-1}\mathbf
{X}_{\mathbf{y}}^{\t}\bolds{\Sigma}_{\mathbf{y}}^{-1}\mathbf{y}$
is the generalized least squares estimator of~$\bm{\beta}$. The associated
mean-squared prediction error (kriging variance) of this predictor
is given by
%
%e2 #&#
\begin{eqnarray}
\label{equk-variance} \sigma_{k}^{2}(\mathbf{s}_{0},t_{0})
& =&\sigma_{0}^{2}-\mathbf {c}_{0\mathbf{y}}^{\t}
\Sigma_{\mathbf{y}}^{-1}\mathbf{c}_{0\mathbf
{y}}
\nonumber
\\[-8pt]
\\[-8pt]
\nonumber
&&{} +\bigl(\mathbf{x}_{0}-\mathbf{c}_{0\mathbf{y}}^{\t}
\bolds{\Sigma }_{\mathbf{y}}^{-1}\mathbf{X}_{\mathbf{y}}\bigr)
\bigl(\mathbf{X}_{\mathbf{y}}^{\t
}\bolds{\Sigma}_{\mathbf{y}}^{-1}
\mathbf{X}_{\mathbf{y}}\bigr)^{-1}\bigl(\mathbf {x}_{0}-
\mathbf{c}_{0\mathbf{y}}^{\t}\bolds{\Sigma}_{\mathbf
{y}}^{-1}
\mathbf{X}_{\mathbf{y}}\bigr)^{\t},
\end{eqnarray}
where $\sigma_{0}^{2}=\operatorname{Var}(Z(\mathbf{s}_{0}))$ [\citet
{Schabenberger2005Statistical}, page~242].
Using the assumptions of normality made regarding $Z(\cdot,\cdot)$
and $\varepsilon(\cdot,\cdot)$, and using the properties of $\hat
{Z}(\cdot,\cdot)$
and $\sigma_{k}(\cdot,\cdot)$, the quantity
%
%e3 #&#
\begin{equation}
\frac{\hat{Z}(\mathbf{s}_{0},t_{0})-Z(\mathbf{s}_{0},t_{0})}{\sigma
_{k}(\mathbf{s}_{0},t_{0})}\label{eqpivotal-quantity}
\end{equation}
has a standard Gaussian distribution. Assuming the null hypothesis
is true, \eqref{eqpivotal-quantity}~leads us to the natural test
statistic
\[
Z^{\prime}(\mathbf{s}_{0},t_{0})=\frac{\hat{Z}(\mathbf
{s}_{0},t_{0})-u}{\sigma_{k}(\mathbf{s}_{0},t_{0})}
\]
for location $\mathbf{s}_{0}$ and time $t_{0}$.\vadjust{\goodbreak}

The confidence level of our confidence region is controlled by carefully
using the duality that exists between confidence intervals and hypothesis
testing. Our confidence region $S_{u^{+}}$ is composed of all locations
$\mathbf{s}$ where we fail to reject the null hypothesis that $Z(\mathbf
{s},t_{p})=u$
[or perhaps more clearly, where we fail to reject that $Z(\mathbf
{s},t_{p})\geq u$].
Thus, our confidence region will fail to contain the true exceedance
region any time a Type I error is made in our hypothesis tests. In
order to maintain a confidence level of $1-\alpha$ for our confidence
region $S_{u^{+}}$, we must ensure that the probability of making
a Type I error for all tests \emph{simultaneously} is~$\alpha$. Let
$C_{\alpha}$ be the critical value at which we will reject $H_{0}$
and conclude $H_{a}.$ By definition, we can only make a Type I error
at the locations $\mathbf{s}$ that are part of the true exceedance
region. Consequently, when seeking to control our error rate, we are
only concerned with controlling our error rate in the context of the
possible realizations of $E_{u^{+}}^{Z}$ conditional on the observed
data. Thus, to control the Type I error rate of our hypothesis at
level $\alpha$ and maintain the confidence level of our confidence
region at $1-\alpha$, we should only conclude $H_{a}$ when
$Z^{\prime}(\mathbf{s},t_{p})<C_{\alpha}$,
where $C_{\alpha}$ is chosen so that
%
%e4 #&#
\begin{equation}
P \Bigl(\inf_{\mathbf{s}\in E_{u^{+}}^{Z}}\bigl\{Z^{\prime}(\mathbf{s},t_{p})
\bigr\} <C_{\alpha} \big|\mathbf{y} \Bigr)=\alpha.\label{eqcv}
\end{equation}

%s2.3 #&#
\subsection{\texorpdfstring{Estimating the critical value {$C_{\alpha}$} and constructing {$S_{u^{+}}$}}
{Estimating the critical value C alpha and constructing S u+}}\label{subEstimating-the-Critical-Value}

To properly estimate the critical value $C_{\alpha}$, we must be able
to adequately approximate the distribution of
%
%e5 #&#
\begin{equation}
\inf_{\mathbf{s}\in E_{u^{+}}^{Z}}\bigl\{Z^{\prime}(\mathbf{s},t_{p})|
\mathbf{y}\bigr\} \label{eqcv-random}
\end{equation}
within our domain $D$. Note that the randomness of \eqref{eqcv-random}
is driven by the uncertainty in $E_{u^{+}}^{Z}$, not the uncertainty
in $\{Z^{\prime}(\mathbf{s},t_{p})\}$. Conditional on the observed data,
$\{Z^{\prime}(\mathbf{s},t_{p})\}$ is completely determined. In contrast,
we do not know the exceedance region $E_{u^{+}}^{Z}$ generated in
the same realization as the observed data $\mathbf{y}$. Though we
are interested in the realization of $E_{u^{+}}^{Z}$ generated along
with the actual data, there are infinitely many possible exceedance
regions compatible with the observed data. Thus, we need to consider
the distribution of $\inf_{\mathbf{s}\in E_{u^{+}}^{Z}}\{Z^{\prime}(\mathbf
{s},t_{p})\}$
for all possible exceedance region realizations compatible with the
observed data.

In practice, to estimate $C_{\alpha}$ in \eqref{eqcv}, we must assume
that the behavior of the hidden process $Z(\cdot,t_{p})$ over the
continuous domain $D$ can be adequately approximated by considering
its behavior over an appropriately sized finite regular grid [cf.
\citet{Zhang2008Loss}]. We discretize the domain $D$ into $m$
pixels, letting $G=\{g_{1},\ldots,g_{m}\}$ be the set of pixels and
$\mathbf{s}_{G}=\{\mathbf{s}_{1}^{*},\ldots,\mathbf{s}_{m}^{*}\}$
be the set of midpoints of these pixels. Let $\mathbf{z}_{G}=[Z(\mathbf
{s}_{1}^{*},t_{p}),\ldots,Z(\mathbf{s}_{m}^{*},t_{p})]^{\t}$
denote the vector of hidden responses on the grid of locations $\mathbf{s}_{G}$.
To approximate possible exceedance region realizations conditional
on the observed data, we can simulate realizations of $\mathbf
{z}_{G}|\mathbf{y}$
and determine the (discretized) exceedance regions for each realization.

The joint distribution of $\mathbf{z}_{G}$ and $\mathbf{y}$ is multivariate
normal, making it straightforward to determine that the conditional
distribution $\mathbf{z}_{G}|\mathbf{y}$ is also multivariate normal
with closed form expressions for the associated mean and covariance
matrix$ $ [\citet{Wasserman2004AllofStatistics}, Theorem 2.44].
It is easy to simulate directly from a multivariate normal distribution
using a decomposition of its covariance matrix [\citet
{Givens2005ComputationalStatistics}, page~146],
but this does not take into account the fact that our mean function
is being estimated. Instead, we will generate realizations of $\mathbf
{z}_{G}|\mathbf{y}$
by conditioning our simulation using kriging. In essence, this approach
produces realizations of the conditional distribution by perturbing
the kriging estimates by possible realizations of the associated kriging
error, as described below.

Suppose that
%
%e6 #&#
\begin{equation}
\left[ \matrix{ \mathbf{y}_{c}
\vspace*{2pt}\cr
\mathbf{z}_{c} }
\right]\sim N\biggl( [
\mathbf{0} ],\left[ \matrix{
\bolds{
\Sigma}_{\mathbf{y}} & \bolds{\Sigma}_{G\mathbf{y}}^{\t}
\vspace*{2pt}\cr
\bolds{\Sigma}_{G\mathbf{y}} & \bolds{\Sigma}_{G} }
\right] \biggr),\label{eqyc-and-zc-distribution}
\end{equation}
recalling that $\bm{\Sigma}_{\mathbf{y}}$ is the $n\times n$ covariance
matrix of the observed data, and letting $\bm{\Sigma}_{G}$ denote
the $m\times m$ covariance matrix of $\mathbf{z}_{G}$ and $\bm{\Sigma
}_{G\mathbf{y}}$
denote the $m\times n$ cross-covariance matrix between $\mathbf{y}$
and $\mathbf{z}_{G}$. The joint distribution of $\mathbf{y}_{c}$
and $\mathbf{z}_{c}$ has the same covariance as the joint distribution
of $\mathbf{y}$ and $\mathbf{z}_{G}$. To obtain $B$ realizations
of $\mathbf{z}_{G}|\mathbf{y}$, we first use an extension of \eqref
{equk-prediction}
allowing for simultaneous predictions at multiple locations [\citet
{Schabenberger2005Statistical}, pages~242--243]
to find the $n\times m$ weight matrix $\bm{\Lambda}$ such that $\hat
{\mathbf{z}}_{G}=[\hat{Z}(\mathbf{s}_{1}^{*},t_{p}),\ldots,\hat
{Z}(\mathbf{s}_{m},t_{p})]^{\t}=\bm{\Lambda}^{\t}\mathbf{y}$.
Next, we simulate $B$ realizations from the distribution in \eqref
{eqyc-and-zc-distribution}.
Denote the first $n$ elements of the $i$th realization by $\mathbf{y}_{c}^{(i)}$
and the next $m$ elements by $\mathbf{z}_{c}^{(i)}$. The $i$th
simulated realization of $\mathbf{z}_{G}|\mathbf{y}$ is then obtained
using the expression
%
%e7 #&#
\begin{equation}
\tilde{\mathbf{z}}^{(i)}=\hat{\mathbf{z}}_{G}+\bigl(
\mathbf{z}_{c}^{(i)}-\bm {\Lambda}^{\t}
\mathbf{y}_{c}^{(i)}\bigr),\label{eqconditional-sim}
\end{equation}
with $[\tilde{Z}^{(i)}(\mathbf{s}_{1}^{*},t_{p}),\ldots,\tilde
{Z}^{(i)}(\mathbf{s}_{m}^{*},t_{p})]^{\t}$
denoting the individual responses of this realization. The first part
of expression \eqref{eqconditional-sim} is the kriging predictor
and the second part is a realization of possible kriging error. Additional
information about this simulation process can be found in \citeauthor{Chiles1999Geostatistics}
[(\citeyear{Chiles1999Geostatistics}), pages~465--468].

To estimate $C_{\alpha}$, we begin by generating $B$ realizations
$\{\tilde{\mathbf{z}}^{(1)},\ldots,\tilde{\mathbf{z}}^{(B)}\}$ of
the conditional process $\mathbf{z}_{G}|\mathbf{y}$. For realization
$i$, we identify $E_{u^{+}}^{\tilde{Z}^{(i)}}=\{\cup g_{j}\dvtx\break\tilde{Z}^{(i)}(\mathbf{s}_{j}^{*},t_{p})\geq u\}$,\vspace*{-1pt}
the union of the pixels where the simulated discretized conditional
process exceeds the threshold $u$. For each realization of
$E_{u^{+}}^{\tilde{Z}^{(i)}}$,
we next consider the minimum of the test statistic $Z^{\prime}(\cdot,t_{p})$
at the gridded locations in the realized exceedance region under the
assumption that the null hypothesis is true. Specifically, for each
realization of $E_{u^{+}}^{\tilde{Z}^{(i)}}$ we find
\[
\min_{\mathbf{s}\in\{\mathbf{s}_{G}\in E_{u^{+}}^{\tilde{Z}^{(i)}}\}}\bigl\{ Z^{\prime}(\mathbf{s},t_{p})
\bigr\}=\min_{\mathbf{s}\in\{\mathbf{s}_{G}\in
E_{u^{+}}^{\tilde{Z}^{(i)}}\}} \biggl\{\frac{\hat{Z}(\mathbf
{s},t_{p})-u}{\sigma_{k}(\mathbf{s},t_{p})} \biggr\}.
\]
Last, $\hat{C}_{\alpha}$, our estimated value of $C_{\alpha}$,
can be obtained by finding the $\alpha$ quantile of the set of minima
from the previous step. The confidence region $S_{u^{+}}$ (or at
least its discretized version) is then $S_{u^{+}}=\{\cup g_{i}\dvtx
Z^{\prime}(\mathbf{s}_{i}^{*},t_{p})\geq\hat{C}_{\alpha}\}$.

%s2.4 #&#
\subsection{Details of inference}

We briefly summarize the inferences that can be made using the proposed
methodology, as well as some details related to this inference.

%s2.4.1 #&#
\subsubsection{Confidence region above}\label{subConfidence-Region-Above}

Using the methodology described above, one may construct a confidence
region $S_{u^{+}}$ containing the realized exceedance region $E_{u^{+}}^{Z}$
with probability $1-\alpha$. Specifically, the probability that the
confidence region $S_{u^{+}}$ produced by our method will contain
the realized exceedance region $E_{u^{+}}^{Z}$ is $1-\alpha$, that is,
%
%e8 #&#
\begin{equation}
P\bigl(E_{u^{+}}^{Z}\subseteq S_{u^{+}}\bigr)=1-\alpha.
\label{eqconf-statement-1}
\end{equation}

Note that the confidence properties of $S_{u^{+}}$ are in the frequentist
paradigm. Thus, the probability in \eqref{eqconf-statement-1} may
refer to the probability of $S_{u^{+}}$ containing the realized exceedance
region before we collect the data or the long-term relative frequency
of containing the realized exceedance region when applying this methodology
to independent trials. A specific confidence region will either contain
or not contain the true but unobserved realization of $E_{u^{+}}^{Z}$,
but in repeated application of this methodology to new data sets,
the long-term proportion of trials where the realized exceedance region
will be entirely contained in the confidence region is $1-\alpha$.

%s2.4.2 #&#
\subsubsection{Confidence region below}

One may obtain a confidence region\break $S_{u^{-}}$ for the exceedance
region below a threshold, $E_{u^{-}}^{Z}$, by considering the inverse
of the~problem described in Sections \ref{subMethodology-for-Calpha}
and \ref{subEstimating-the-Critical-Value}. By testing for all\break $\mathbf
{s}\in D$,
$H_{0}\dvtx Z(\mathbf{s},t_{p})\leq u$ versus $H_{a}\dvtx Z(\mathbf{s},t_{p})>u$
[and modifying the steps in~Section~\ref{subEstimating-the-Critical-Value}
so that we approximate the critical value $C_{\alpha}$ such that\break
$P(\sup_{\mathbf{s}\in E_{u^{-}}^{Z}}\{Z{}^{\prime}(\mathbf{s},t_{p})\}
>C_{\alpha}|\mathbf{y})=\alpha$],
we can construct a confidence region $S_{u^{-}}$ for $E_{u^{-}}^{Z}$
by letting $S_{u^{-}}=\{\mathbf{s}\in D\dvtx Z^{\prime}(\mathbf
{s},t_{p})\leq C_{\alpha}\}$.
This type of inference will be useful when one desires a confidence
region for the entire area where a process may fall below some level,
for example, identifying all locations where drought may occur. The confidence
properties of $S_{u^{-}}$ are the same as those described for $S_{u^{+}}$
in Section \ref{subConfidence-Region-Above}.

%s2.4.3 #&#
\subsubsection{Properties of the complement}

The inference related to $S_{u^{+}}$ and $S_{u^{-}}$ is useful when
one desires to find a region containing an entire exceedance region
with high confidence. However, it is possible that a sizable portion
of the confidence region is not part of the exceedance region and
that the confidence region is much larger than the true exceedance
region. Instead, we may be interested in the region of our domain
$D$ where we can be confident that a response will exceed the threshold
in question. This type of confidence region is easily obtained as
a byproduct of the construction of $S_{u^{+}}$ and $S_{u^{-}}$.

Suppose we wish to find a confidence set contained entirely in $E_{u^{+}}^{Z}$,
that is, every point in this confidence set has a response exceeding
the threshold $u$. When constructing the confidence region $S_{u^{-}}$
for the locations having a response below the threshold $u$, $S_{u^{-}}$
has the property that $P(E_{u^{-}}^{Z}\subseteq S_{u^{-}})=1-\alpha$.
Letting $S_{u^{-}}^{c}$ denote the complement of $S_{u^{-}}$, this
implies that $P(\exists\mathbf{s}\in S_{u^{-}}^{c}\dvtx Z(\mathbf
{s},t_{p})\leq u)=\alpha$,
or, more naturally, $P(\forall\mathbf{s}\in S_{u^{-}}^{c}\dvtx Z(\mathbf
{s},t_{p})>u)=1-\alpha$,
which implies that
\[
P\bigl(S_{u^{-}}^{c}\subseteq E_{u^{+}}^{Z}
\bigr)=1-\alpha.
\]
In other words, we can be confident that every point in $S_{u^{-}}^{c}$
simultaneously has a response exceeding the threshold $u$. Similar
to the properties of the rejection region $S_{u^{-}}^{c}$, the rejection
region $S_{u^{+}}^{c}$ has the property that every point in it will
fall below the threshold $u$ with confidence.

By combining the information from $S_{u^{-}}^{c}$ with $S_{u^{+}}$,
one may obtain a liberal and conservative view of where the realized
exceedance region $E_{u^{+}}^{Z}$ may be located. Specifically, we
can be confident (in a frequentist sense) that every point in $S_{u^{-}}^{c}$
is a member of the true exceedance set $E_{u^{+}}^{Z}$, but it is
likely that $S_{u^{-}}^{c}$ is smaller than the true exceedance set.
On the other hand, we can be confident that every location in $E_{u^{+}}^{Z}$
will be contained within $S_{u^{+}}$, but $S_{u^{+}}$ will likely
be larger than $E_{u^{+}}^{Z}$. A similar interpretation applies
to the regions $S_{u^{+}}^{c}$ and $S_{u^{-}}$ in relation to $E_{u^{-}}^{Z}$.
Table \ref{tabSummary-of-procedure} summarizes the methodology and
inferences resulting from the proposed methodology.

%t1 #&#
\begin{table}
\tabcolsep=4pt
\caption{Summary of procedure and inferences for proposed methodology
\label{tabSummary-of-procedure}}
\begin{tabular*}{\textwidth}{@{\extracolsep{\fill}}lcc@{}}
\hline
\multicolumn{1}{c}{} & \multicolumn{1}{c}{\textbf{Exceedance above}} &
\multicolumn{1}{c@{}}{\textbf{Exceedance below}}\\
\hline
Region & $E_{u^{+}}^{Z}=\{\mathbf{s}\in D\dvtx Z(\mathbf{s},t_{p})\geq
u\}$ & $E_{u^{-}}^{Z}=\{\mathbf{s}\in D\dvtx Z(\mathbf{s},t_{p})\leq u\}
$\\
$H_{0}$ & $Z(\mathbf{s},t_{p})\geq u$ & $Z(\mathbf{s},t_{p})\leq
u$\\
$H_{a}$ & $Z(\mathbf{s},t_{p})<u$ & $Z(\mathbf
{s},t_{p})>u$\\[2pt]
Test stat. & \multicolumn{2}{c}{$Z^{\prime}(\mathbf{s},t_{p})={ \frac
{Z(\mathbf{s},t_{p})-u}{\sigma_{k}(\mathbf{s},t_{p})}}$}\\[3pt]
Crit. value $C_{\alpha}$& $P(\inf_{\mathbf{s}\in E_{u^{+}}^{Z}}\{Z^{\prime}(\mathbf
{s},t_{p})\}<C_{\alpha}|\mathbf{y}) =\alpha$ & $P(\sup_{\mathbf{s}\in
E_{u^{-}}^{Z}}\{Z^{\prime}(\mathbf{s},t_{p})\}>C_{\alpha}|\mathbf
{y})=\alpha$\\
Conf. region & $S_{u^{+}}=\{\mathbf{s}\in D\dvtx Z^{\prime}(\mathbf
{s},t_{p})\geq C_{\alpha}\}$ & $S_{u^{-}}=\{\mathbf{s}\in D\dvtx
Z^{\prime}(\mathbf{s},t_{p})\leq C_{\alpha}\}$\\
Inference 1 & $P(E_{u^{+}}^{Z}\subseteq S_{u^{+}})=1-\alpha$ &
$P(E_{u^{-}}^{Z}\subseteq S_{u^{-}})=1-\alpha$ \\
Inference 2 & $P(S_{u^{+}}^{c}\subseteq E_{u^{-}}^{Z})=1-\alpha$ &
$P(S_{u^{-}}^{c}\subseteq E_{u^{+}}^{Z})=1-\alpha$\\
\hline
\end{tabular*}
\end{table}

%s2.4.4 #&#
\subsubsection{Additional details}\label{subAdditional-Details}

We note once again that the inference and properties of confidence
regions discussed above are directly related to the domain in question.
Since the procedure described in Section \ref{subEstimating-the-Critical-Value}
depends on discretizing the domain of interest $D$, determining the
realized exceedance regions $E_{u^{+}}^{\tilde{Z}^{(i)}}$ over that
domain, estimating $C_{\alpha}$ over that domain, and evaluating
the hypothesis tests over that domain, the confidence regions are
directly linked to the domain chosen. This must be kept in mind when
deciding on a domain of interest and considering the scope of inference.

We also note that the confidence properties above are dependent on
knowing the true covariance properties of the random field under consideration.
In practice, this is rarely the case. While it is typical to simply
plug-in the estimated covariance parameters as if they were the truth
[\citet{Schabenberger2005Statistical}, page~254], this will likely
result in confidence regions that have confidence levels differing
from the desired levels. A Bayesian approach to this problem that
naturally incorporates the uncertainty in the covariance parameters
is currently under development.

%s3 #&#
\section{Simulation studies}\label{secSimulation-Studies}

In this section we will use simulation experiments of the proposed
methodology to compare empirical confidence levels to intended confidence
levels for scenarios involving several combinations of mean structures
and spatio-temporal dependence. In the experiments, the goal is to
construct a confidence region $S_{u^{+}}$ containing the exceedance
region $E_{u^{+}}^{Z}$ of the realization in question. The experiments
are similar to ones performed by \citet{Zhang2008Loss}. Discussion
will follow regarding how estimating the covariance parameters affects
the empirical confidence levels and how the covariance parameters
affect the shape and size of the confidence region for the true exceedance
set. The analysis in Sections \ref{secSimulation-Studies}, \ref
{secApplication-1},
and \ref{secApplication-2} was performed using R version 2.15.1
[R Core Team (\citeyear{R2151})].

%s3.1 #&#
\subsection{Overall structure}\label{subOverall-Structure}

\begin{figure}
\centering
\begin{tabular}{@{}cccc@{}}
\footnotesize{(a) Trend} & \footnotesize{(b) Cone} & \footnotesize{(c)
Cup}& \footnotesize{(d) Waves}\\

\includegraphics{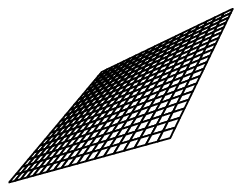}
 & \includegraphics{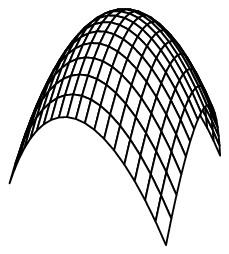} & \includegraphics{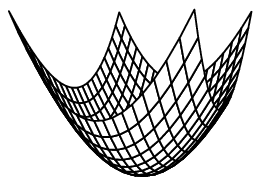}& \includegraphics{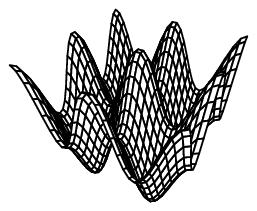}
\end{tabular}
\caption{Mean structure of four different patterns.}\label
{figMean-structure-of-patterns}
\end{figure}

In each of the subsequent experiments, the hidden process $Z(\cdot,\cdot)$
was assumed to follow the same general form
%
%e9 #&#
\begin{equation}
Z(\mathbf{s},t)=\mathbf{x}(\mathbf{s},t)^{\t}\bm{\beta }+W(
\mathbf{s},t).\label{eqmodel-form}
\end{equation}
The mean of $Z(\mathbf{s},t)$ was allowed to vary between four different
patterns: trend, cone, cup, and waves. The domain $D$ of the experiment
depended on the pattern under consideration. The mean structures of
these patterns are shown in Figure \ref{figMean-structure-of-patterns},
while the corresponding domain $D$ of the experiment, the covariate
vector $\mathbf{x}(\mathbf{s},t)$, and coefficient vector $\bm{\beta}$
of each pattern are provided in Table \ref{tabexperimental-parameters}.

%f1 #&#

%t2 #&#
\begin{table}[b]
\caption{The domain $D$, covariates vector $\mathbf{x}(\mathbf{s},t)$, and
coefficient vector $\bm{\beta}$ used for the experiments of each
pattern shown in Figure \protect\ref{figMean-structure-of-patterns}. The
individual spatial coordinates of a spatial location $\mathbf{s}$
are denoted $\mathbf{s}_{x}$ and $\mathbf{s}_{y}$}\label
{tabexperimental-parameters}
\begin{tabular*}{\textwidth}{@{\extracolsep{\fill}}lccc@{}}
\hline
\multicolumn{1}{@{}l}{\textbf{Pattern}} & $\bolds{D}$ & $\mathbf{x}\bolds{(}\mathbf{s},\bolds{t}\bolds{)}$ &
\multicolumn{1}{c@{}}{$\bm{\beta}$}\\
\hline
Trend & $[0,1]\times[0,1]$ & $[1,\mathbf{s}_{x},\mathbf{s}_{x}]^{\t}$ &
$ [1,3,3]^{\t}$\tabularnewline
Cone & $[-0.5,0.5]\times[-0.5,0.5]$ & $[1,\mathbf{s}_{x}^{2},\mathbf
{s}_{y}^{2}]^{\t}$ & $[1,-20,-20]^{\t}$\tabularnewline
Cup & $[-0.5,0.5]\times[-0.5,0.5]$ & $[1,\mathbf{s}_{x}^{2},\mathbf
{s}_{y}^{2}]^{\t}$ & $[1,20,20]^{\t}$\tabularnewline
Waves & $[-\frac{3}{2}\pi,\frac{5}{2}\pi]\times[-2\pi,2\pi]$ & $[1,\cos
(\mathbf{s}_{x}),\sin(\mathbf{s}_{y})]^{\t}$ & $[1,5,5]^{\t
}$\\
\hline
\end{tabular*}
\end{table}
The spatio-temporal process $W(\mathbf{s},t)$ was multivariate normal
with mean 0 and isotropic covariance function
%
%e10 #&#
\begin{equation}
C(\mathbf{s}_{1}-\mathbf{s}_{2},t_{1}-t_{2})=
\sigma_{W}^{2}\exp\bigl(\Vert \mathbf{s}_{1}-
\mathbf{s}_{2}\Vert/\phi\bigr)\rho^{|t_{1}-t_{2}|},\label {eqcorrelation-structure}
\end{equation}
where $\Vert\cdot\Vert$ is the Euclidean distance metric. This covariance
function is obtained by multiplying an exponential covariance function
$\exp(\Vert\mathbf{s}_{1}-\mathbf{s}_{2}\Vert/\phi)$ by $\rho
^{|t_{1}-t_{2}|},$
the correlation function of an $\operatorname{AR}(1)$ temporal process. The strength
of the spatial dependence is determined by $\phi$, and the strength
of the temporal dependence is governed by $\rho$. In all experiments,
the scale parameter $\sigma_{W}^{2}$ was fixed to be 1, while the
spatial dependence parameter $\phi$ varied between the values $0.5$
(weak spatial dependence), $1.5$ (moderate spatial dependence), and
5 (very strong spatial dependence), and the temporal dependence parameter
$\rho$ varied between the values~$0.1$ (weak temporal dependence),
$0.5$ (moderate temporal dependence), and 0.9 (strong temporal dependence).
Experiments were performed assuming no measurement error ($\sigma
_{\varepsilon}^{2}=0$)
and when measurement error was present ($\sigma_{\varepsilon}^{2}=0.1$
or 0.5 depending on the mean structure).

In each experiment, the temporal window of interest $T=\{1,2,3,4\}$.
It was assumed that responses with time index $t=1,2,3$ were observed
data and that responses with time index $t=4$ corresponded to future
responses. For each of the first three times, responses were observed
at the same $100$ locations $\mathbf{s}_{1},\mathbf{s}_{2},\ldots,\mathbf{s}_{100}$.
The sites of the 100 locations were irregularly spaced and were obtained
by drawing independent observations from a $\operatorname{Uniform}(0,1)$ distribution
and combining them to form a set of spatial coordinates. To apply
the methodology described in Section \ref{secMethodology} and obtain
``future'' data to test empirical confidence levels, the spatial
domain for each experiment was discretized into a $50\times50$ grid
of pixels. For the trend pattern with domain $D=[0,1]\times[0,1]$,
this resulted in the set of pixels $G=\{[x,x+0.02]\times[y,y+0.02]\dvtx
x,y\in\{0,0.02,\ldots,0.98\}\}$,
and the set of pixel center points $\mathbf{s}_{G}=\{(x,y)\dvtx x,y\in\{
0.01,0.03,\ldots,0.99\}\}$.
In the fourth year, future responses were ``observed'' at all locations
$\mathbf{s}\in\mathbf{s}_{G}$ to provide the future test data. For
each combination of covariance parameters $\phi$ and $\rho$, the
hidden process $Z(\cdot,\cdot)$ was randomly generated at each of
the 100 locations for the first three times, and then at the pixel
center points for the fourth time. After generating the hidden process
$Z(\cdot,\cdot)$, independent error values $\varepsilon_{1},\varepsilon
_{2},\ldots,\varepsilon_{300}$
were generated according to a $N(0,\sigma_{\varepsilon}^{2})$ distribution
and added to the 300 hidden process values observed for the first
three times to obtain the observed response values $\mathbf{y}=\{
Z(\mathbf{s}_{1},1)+\varepsilon_{1},\ldots,Z(\mathbf
{s}_{100},1)+\varepsilon_{100},Z(\mathbf{s}_{1},2)+\varepsilon
_{101},\ldots,Z(\mathbf{s}_{100},2)+\varepsilon_{200},Z(\mathbf
{s}_{1},3)+\varepsilon_{201},\ldots,Z(\mathbf{s}_{100},3)+\varepsilon
_{300}\}$.
Note that the observed process recovers the hidden process when $\sigma
_{\varepsilon}^{2}=0$.

The threshold level $u$ used to create the exceedance set $E_{u^{+}}^{Z}$
was the 90th percentile of the data generated on the $50\times50$
grid for time $t=4$. Confidence regions $S_{u^{+}}$ were constructed
at confidence levels of 0.90 and 0.95. For each experiment, the critical
value $C_{\alpha}$ was estimated using 2000 realizations of $\mathbf
{z}_{G}|\mathbf{y}$.
Empirical confidence levels were calculated by generating 200 independent
realizations of $Z(\cdot,\cdot)$ for each experimental setting, using
the procedure of Section \ref{secMethodology} to construct the confidence
region $S_{u^{+}}$ for the exceedance region of the hidden process
at time $t=4$, and then finding the proportion of realizations in
which the confidence region $S_{u^{+}}$ contained the exceedance
region $E_{u^{+}}^{Z}$.

%s3.2 #&#
\subsection{Empirical confidence levels}\label{subEmpirical-Confidence-Levels}

For the trend pattern, experiments were run for all 9 combinations
of the two dependence parameters $\phi$ and $\rho$. Experiments
were run both without measurement error ($\sigma_{\varepsilon}^{2}=0$)
and with measurement error ($\sigma^{2}=0.1$). In the initial experiments,
it was assumed that the covariance parameters were all known. Subsequently,
the same experiments were performed assuming that the covariance parameters
were unknown. In the second set of experiments, the covariance parameters
were estimated using restricted maximum likelihood (REML). The measurement
error variance $\sigma_{\varepsilon}^{2}$ was estimated in each of these
experiments, regardless of whether measurement error was actually
present.

%s3.2.1 #&#
\subsubsection{Results when covariance parameters known}\label
{subResults-when-covariance-known}

\begin{figure}[b]
\centering
\begin{tabular}{@{}cc@{}}

\includegraphics{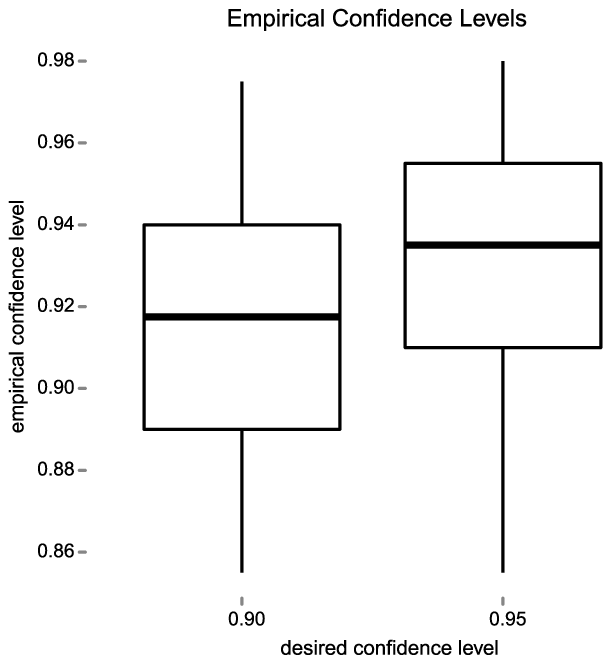}
 & \includegraphics{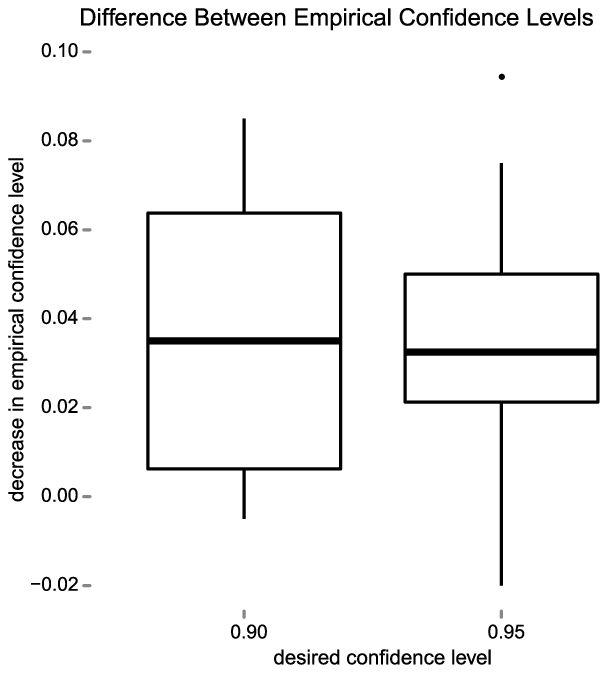}
\\
\footnotesize{(a)} & \footnotesize{(b)}
\end{tabular}
\caption{Boxplots of the empirical confidence levels for the confidence regions
produced by the simulation study are shown by confidence level in
\textup{(a)}. The change in empirical confidence level when estimated covariance
parameters were used to construct the confidence regions is shown
in \textup{(b)}. \label{figBoxplots-of-the-empirical-conf-levels}}
\end{figure}

The results of the simulation studies for the trend mean structure
when the covariance parameters were assumed known are given below
in Table \ref{tabsim-results-no-error}. A comparative boxplot of
the empirical confidence levels for these experiments grouped by desired
confidence level is shown in Figure \ref
{figBoxplots-of-the-empirical-conf-levels}(a). The empirical confidence levels are all fairly close to the desired
confidence level, and the empirical confidence levels behave in a
manner consistent with sampling variability.

%t3 #&#
\begin{table}
\caption{Empirical coverage rate of confidence region procedure for trends
simulation study. The standard error of the estimates are 2.12\% for
the 0.90 confidence level and 1.54\% for the 0.95 confidence level}\label{tabsim-results-no-error}
\begin{tabular*}{\textwidth}{@{\extracolsep{\fill}}lccccccccc@{}}
\hline
& \multicolumn{1}{r}{} & \textbf{Error variance} & \multicolumn{1}{c}{$\bolds{\sigma
_{\varepsilon}^{2}}$} & \multicolumn{3}{c}{\textbf{0}} & \multicolumn
{3}{c@{}}{\textbf{0.1}}\\[-6pt]
&  &  &  & \multicolumn{3}{c}{\hrulefill} & \multicolumn
{3}{c@{}}{\hrulefill}\\
& \multicolumn{1}{r}{} & \textbf{Time dependence} & \multicolumn{1}{c}{$\bolds{\rho}$}
& \textbf{0.1} & \textbf{0.5} & \textbf{0.9} & \textbf{0.1} & \textbf{0.5} &
\multicolumn{1}{c@{}}{\textbf{0.9}}\\
\hline
\multirow{6}{*}{\rotatebox{-90}{Conf. level}} & 0.90& {Spat. depend. $\phi$}
& 0.5 & 0.905 &0.855 & 0.885 &0.900 & 0.880 & 0.875\\
&  & & 1.5 & 0.910 & 0.915 & 0.905 &0.870 & 0.890 &0.860\\
& & & 5\phantom{0.} & 0.885 & 0.910 & 0.890 &0.870 & 0.915 & 0.920\\[3pt]
&0.95 & Spat. depend. $\phi$& 0.5 & 0.960 & 0.915 & 0.960 & 0.940 & 0.940 &0.930\\
&  &   & 1.5 & 0.965 & 0.970 & 0.940 & 0.930 &0.94\phantom{0} & 0.925\\
& & & 5\phantom{0.} & 0.970 & 0.950 & 0.940 & 0.930 & 0.935 & 0.975\\
\hline
\end{tabular*}
\end{table}

%f2 #&#

%s3.2.2 #&#
\subsubsection{Results when covariance parameters unknown}

The covariance parameters of our data will not be known in practice,
so we must use estimated values when making predictions. The empirical
confidence level of the confidence regions of Section \ref{secMethodology}
are likely to be lower when estimated covariance parameter values
are used instead of the true covariance parameter values. To get an
idea of the magnitude of this drop, the simulation experiments for
the trends mean structure of the previous section were repeated using
estimated covariance parameters. Utilizing the same random number
seed in each set of experiments, the observed data for the two
sets of experiments was the same. The only difference in inference
was that the covariance parameters were estimated via restricted maximum
likelihood (REML) in the second set of experiments and then used in
the construction of the confidence regions. Consequently, the results
from the two experiments are paired. The differences between the empirical
confidence levels for the two sets of experiments are displayed in
the boxplot in Figure \ref{figBoxplots-of-the-empirical-conf-levels}(b). The difference between the empirical confidence level for the
two settings is typically between 0.01--0.07, though the values sometimes
deviated from this range.

%s3.2.3 #&#
\subsubsection{Computational cost}

The proposed methodology has a computational cost essentially
the same as that of the conditional simulation algorithm. As pointed
out in the discussion of \eqref{eqconditional-sim}, the predicted
responses and prediction weights are reused in the conditional simulation
process. The only additional cost is the simulation of the unconditional
random field in \eqref{eqyc-and-zc-distribution}, and this cost
can vary greatly depending on the algorithm selected. For the simulations
described above, the prediction, conditional simulation, and construction
of the confidence regions averaged just under 40 seconds on a MacBook
Pro running OS X 10.7.4 with a 2.53~GHz Intel Core 2 Duo CPU and 4
GB of RAM. The same process averaged just over 3 seconds on a desktop
PC running Fedora Core 16 with a 3.33~GHz Intel Core i7 CPU and 24
GB of RAM since many of the computations can run in parallel. The
memory needed to construct the confidence regions depends directly
on the number of observed and predicted data values and depends on
the implementation of the proposed methodology. The conditional simulation
process can be performed one at a time if necessary to minimize memory
usage. However, for the simulation experiments performed in this section,
the objects directly related to prediction, conditional simulation,
and construction of the confidence regions required just over 38 MB
of RAM.

%s3.3 #&#
\subsection{Shapes}\label{subShapes}

A question of interest is how the shape of the confidence region for
the exceedance set is affected by factors such as the mean structure,
spatio-temporal dependence, and error variance. We briefly studied
the effects of these factors graphically using the cone, cup, and
waves patterns discussed in Section~\ref{subOverall-Structure}.
For these patterns, only the combinations of $(\phi,\rho)$ equal
to (0.1, 0.1), (0.5, 0.5), (0.9, 0.9) were tested; we refer to these
combinations of $(\phi,\rho)$ as weak, medium, and strong spatio-temporal
dependence, respectively. The covariance parameters were assumed known
in these additional experiments. In order to evaluate the the general
shape of the confidence and exceedance regions for each mean structure,
the 200 confidence regions and realized exceedance regions from each
simulation experiment were used to construct a ``median'' exceedance
region and confidence region. The ``median'' in each case was determined
by noting the pixels where the realized exceedance region and/or confidence
region appeared in at least half of the 200 realizations. Similarly,
to evaluate how the strength of spatio-temporal dependence and measurement
error affect the size of a confidence region, the number of pixels
used to construct each confidence region for the 200 realizations
of each experiment was also noted.

%f3 #&#
\begin{figure}
\centering
\begin{tabular}{@{}c@{\hspace*{2pt}}c@{\hspace*{2pt}}c@{}}

\includegraphics{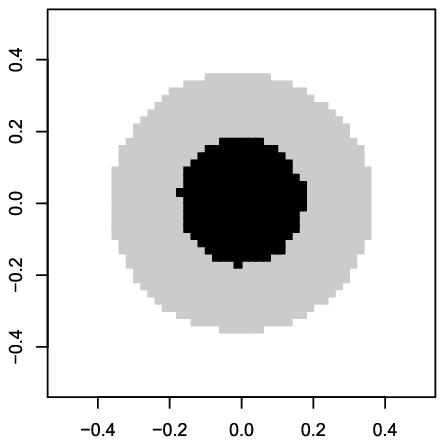}
 & \includegraphics{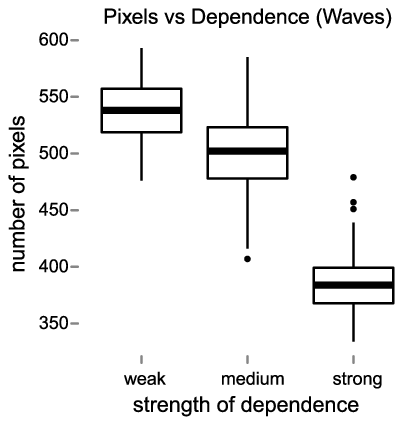} & \includegraphics{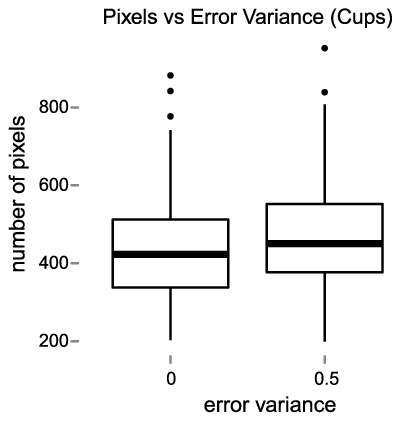}\\
\footnotesize{(a)} & \footnotesize{(b)} & \footnotesize{(c)}
\end{tabular}
\caption{The median confidence region and exceedance set for the cone mean
structure (medium dependence without measurement error) are shown in
\textup{(a)}. Boxplots comparing the size of the confidence region (in number
of pixels) versus the strength of dependence for the waves mean structure
(without measurement error) for each of the 200 simulations are shown
in \textup{(b)}. Boxplots comparing the size of the confidence region (in number
of pixels) versus the measurement error variances for the cups mean
structure (strong dependence) for each of the 200 simulations are shown
in \textup{(c)}.}\label{figMedian-confidence-regions-cone}
\end{figure}

The shapes of the confidence and exceedance regions for the different
mean structures conformed to intuitive expectations. Specifically,
the confidence and exceedance regions of each shape patterned the
mean structure. For example, the confidence and exceedance region
for the cone mean structure were roughly circular since the exceedance
region of the cone would simply be the upper portion of the cone.
As an example of this, the median confidence region and exceedance
set for the cone mean structure having medium spatial and temporal
dependence without measurement error are shown in Figure \ref
{figMedian-confidence-regions-cone}(a). In order to assess the relationship between the strength of spatio-temporal
dependence and the size of the resulting confidence region, as well
as how measurement error affected these regions, we looked at the
number of pixels in the confidence region from each of the 200 realizations
of each experiment. Intuitively, stronger spatio-temporal dependence
would lead to smaller predictive uncertainty, leading to a better
estimate and smaller confidence region for the realized exceedance
region. This pattern was consistently seen across all mean structures,
and an example of this is shown in the boxplots in Figure \ref
{figMedian-confidence-regions-cone}(b) for the waves mean structure. Last, we would expect that the
addition of measurement error to the observed responses would increase
the size of the confidence regions due to greater uncertainty in predictions.
This pattern was seen across all mean structures and levels of dependence,
and an example of this for the cups mean structure is shown in Figure
\ref{figMedian-confidence-regions-cone}(c).

%f4 #&#
\begin{figure}[b]\vspace*{-3pt}
\centering
\begin{tabular}{@{}c@{\hspace*{2pt}}c@{}}

\includegraphics{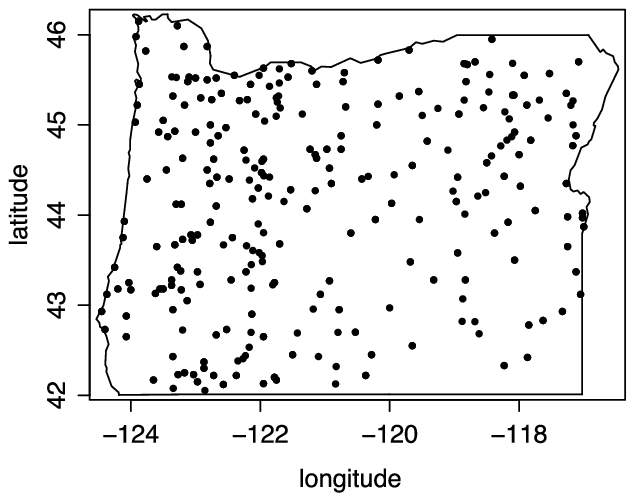}
 & \includegraphics{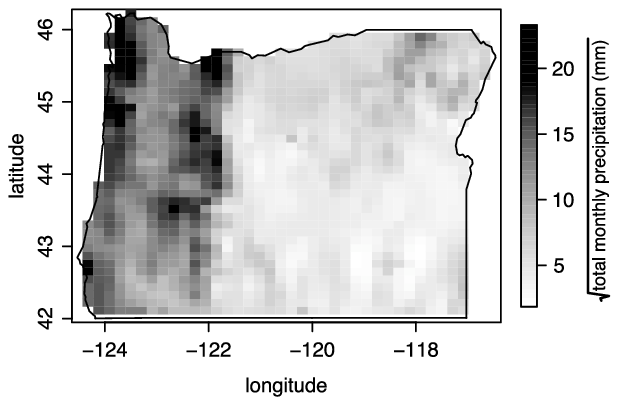}\\
\footnotesize{(a)} & \footnotesize{(b)}
\end{tabular}
\caption{The observed data locations projected on a map of Oregon are shown
in \textup{(a)}. An image plot of the average of the square root of the total
monthly precipitation~(mm) for the months of October 1996 and 1997
for the state of Oregon is shown in \textup{(b)}.}\label{figOregon-exploratory}
\end{figure}

%s4 #&#
\section{Case study 1: Precipitation in Oregon}\label{secApplication-1}

We demonstrate application of this methodology using precipitation
data from the state of Oregon. Our goal is to identify with 90\% confidence
the regions of Oregon where the total monthly precipitation exceeds
250~mm (approximately the 96th percentile of the observed data) in
October of 1998 using the precipitation measurements available in
October of 1996 and 1997. Analysis was performed using the raw data
available at
\texttt{\href{http://www.image.ucar.edu/Data/US.monthly.met/FullData.shtml\#precip}{http://www.image.ucar.edu/Data/US.monthly.met/}\break
\href{http://www.image.ucar.edu/Data/US.monthly.met/FullData.shtml\#precip}{FullData.shtml\#precip}}.
The data used in this case study were created from the data archives
of the National Climatic Data Center and have previously been used
to generate high-resolution maps of precipitation and other meteorological
variables [\citet{Daly2000United,Johns2003Infilling}]. The
raw data provides the longitude (degrees), latitude (degrees), and
elevation~(m) for 11,918 unique sites throughout the United States
and, when available, includes the total monthly precipitation~(mm)
for each month between the years 1895--1997. There were 447 total observations
provided for the state of Oregon between 1996 and 1997, with 191 observations
in 1996 and 256 observations in 1997. The observed data locations
are indicated by solid dots on a map of Oregon in Figure~\ref
{figOregon-exploratory}(a).

%
%}\hfill\subfloat[]{\begin{centering}
%
%}
%
%shown
%in (a). An image plot of the average of the square root of the total
%monthly precipitation (mm) for the months of October 1996 and 1997
%for the state of Oregon is shown in (b). \label{figOregon-exploratory}}

Exploratory analysis of the precipitation measurements revealed them
to be positively skewed, so a square-root transformation of the measurements
was taken to achieve approximate normality. An image plot of average
precipitation for the transformed measurements is shown in Figure
\ref{figOregon-exploratory}(b). Because the data were irregularly
spaced, the bicubic spline interpolation algorithm of \citet
{Akima1996Algorithm}
was used to interpolate the average of the square root of the total
monthly precipitation for the months of October 1996 and 1997 onto
a regular grid before constructing the image plot.

The mean precipitation level of the responses $Z(\mathbf{s},t)$ was
assumed to be the same for each year and follow the linear structure
$\mu(\mathbf{s},t)=\mathbf{x}(\mathbf{s},t)^{\t}\bm{\beta}$,
with the covariates vector $\mathbf{x}(\mathbf{s},t)=[1,\mathrm
{longitude},\mathrm{latitude},\mathrm{elevation}]^{\t}$.
The covariance function of the hidden data was modeled as being stationary
and fully separable with respect to space and time, while allowing
for a spatio-temporal measurement error effect for the observed data
that varied by year. Consequently, the covariance function of the
observed data may be written as $C(\mathbf{s}_{1}-\mathbf
{s}_{2},t_{1}-t_{2})=C_{S}(\mathbf{s}_{1}-\mathbf
{s}_{2})C_{T}(t_{1}-t_{2})+C_{\varepsilon}(\mathbf{s}_{1}-\mathbf
{s}_{2},t_{1}-t_{2})$,
where $C_{S}(\cdot)$ is a purely spatial covariance function,
$C_{T}(\cdot)$
is a purely temporal covariance function, and $C_{\varepsilon}(\cdot,\cdot)$
is the covariance function of the measurement errors. We assumed that
the spatial covariance $C_{S}(\cdot)$ could be modeled using an isotropic,
Mat\'{e}rn covariance model of the form $C_{S}(\mathbf{s}_{1},\mathbf
{s}_{2})=\sigma^{2}2^{1-\nu}(h/\phi)^{\nu}\mathcal{K}_{\nu}(h/\phi
)/\Gamma(\nu)$,
where $h=\Vert\mathbf{s}_{1}-\mathbf{s}_{2}\Vert$ is the Euclidean distance
between two data locations, $\phi$ is a parameter related to the
strength of spatial dependence, $\sigma^{2}$ measures the variance
of the hidden process $Z\st$, $\nu$ controls the smoothness of the
hidden process, and $\mathcal{K}_{\nu}(\cdot)$ is the modified Bessel
function of the second kind of order $\nu$. The temporal covariance
$C_{T}(t_{1},t_{2})$ was modeled using the correlation function of
an $\operatorname{AR}(1)$ process so that $C_{T}(t_{1},t_{2})=\rho^{u}$, where $u=|t_{1}-t_{2}|$
is the time lag between $t_{1}$ and $t_{2}$ and $\rho$ measures
the strength of spatial dependence. Last, the covariance function
of the measurement errors can be written as
\[
C_{\varepsilon}(\mathbf{s}_{1}-\mathbf{s}_{2},t_{1}-t_{2})=
\bigl[\sigma _{96}^{2}\mathbf{1}_{(96)}(t_{1})+
\sigma_{97}^{2}\mathbf {1}_{(97)}(t_{1})
\bigr]\mathbf{1}_{(0,0)}(h,u),
\]
where $\sigma_{96}^{2}$ and $\sigma_{97}^{2}$ are the variances
of the measurement errors in years 1996 and 1997, respectively. Restricted
maximum likelihood estimation was used to estimate the covariance
parameters $\bm{\theta}=[\sigma^{2},\phi,\nu,\rho,\sigma
_{96}^{2},\sigma_{97}^{2}]^{\t}$,
with the resulting estimates being
\[
\hat{\bm{\theta}}=[12.46,1.54,0.53,0.88,0.52,1.01]^{\t}.
\]
Using the estimated covariance parameters to estimate the covariance
matrix of the observed data, the generalized least squares estimates
of the trend parameter vector $\bm{\beta}$ is $\hat{\bm{\beta
}}_{\mathrm{gls}}=[-332.02,-2.17,1.73,0.0039]^{\t}$.
Examining all of the available total monthly precipitation measurements
for the month of October in the state of Oregon from 1895--1997, 250~mm of total rain corresponded to roughly the 96th percentile and was
chosen as the exceedance threshold for which exceedance locations
would be identified in October 1998. Accordingly, on the transformed
scale our goal is to identify at a confidence level of 0.90 the exceedance
region $E_{\sqrt{250}^{+}}^{Z}$ where the transformed total monthly
precipitation measurements exceed $u=\sqrt{250\ \mathit{mm}}$.

%f5 #&#
\begin{figure}

\includegraphics{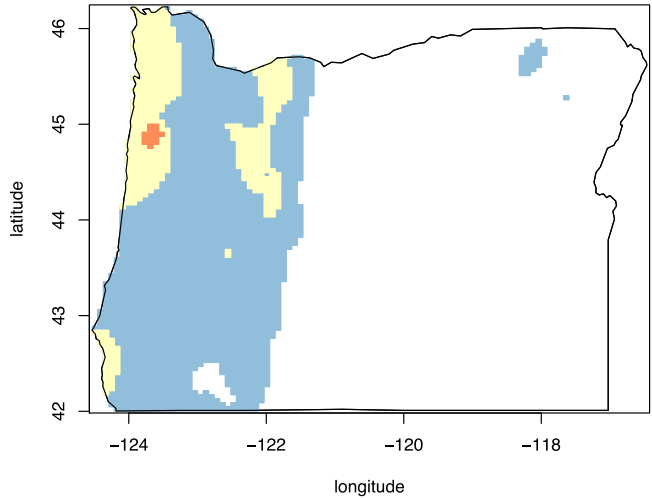}

\caption{The 90 percent confidence regions $S_{\sqrt{250}^{+}}$ and
$S_{\sqrt{250}^{-}}^{c}$
of Oregon where the square root of the total monthly precipitation~(mm) in October of 1998 exceeds $\sqrt{250\ \mathit{mm}}$. The region
$S_{\sqrt{250}^{+}}$ is colored blue, $S_{\sqrt{250}^{-}}^{c}$ is
colored orange, while the region where the predicted values exceed
$\sqrt{250\ \mathit{mm}}$ is colored yellow.}\label{figstate-conf-region}
\end{figure}

The next step in our analysis was creating a discrete grid in the
region of interest~$D$, where $D$ is the state of Oregon. For computational
simplicity, a $100\times100$ rectangular grid was superimposed over
$D$, and the pixels with center points contained in $D$ were used
as the discrete grid in subsequent analysis. Next, the procedure outlined
in Section \ref{subEstimating-the-Critical-Value} was used to construct
both $S_{\sqrt{250}^{+}}$ and $S_{\sqrt{250}^{-}}^{c}$ at a confidence
level of 0.90 using 2000 realizations of \eqref{eqconditional-sim};
the resulting confidence regions are shown in Figure \ref{figstate-conf-region}.
The yellow coloring in Figure \ref{figstate-conf-region} indicates
the regions where the predicted value (using universal kriging) is
greater than the designated threshold, the blue coloring indicates
the 90 percent confidence region $S_{\sqrt{250}^{+}}$, while orange
coloring indicates the 90 percent confidence region $S_{\sqrt{250}^{-}}^{c}$.
Naturally, the confidence region $S_{\sqrt{250}^{+}}$ contains the
predicted exceedance region, while the predicted exceedance region
contains $S_{\sqrt{250}^{-}}^{c}$. The area of Oregon predicted to
have total monthly precipitation greater than 250~mm in October of
1998 is mainly limited to the western area of the state closer to
the coast. The associated 90 percent confidence region $S_{\sqrt{250}^{+}}$
where 250~mm of rain could fall in October of 1998 is also found in
the western portion of the state, though two small regions in the
northeast corner of Oregon were also considered as possible candidates
for this event. Only a small region (shown in orange) in the northwest
part of the state could confidently be predicted to receive at least
250~mm of total monthly precipitation during this time period. On
the other hand, the regions without any shading are the regions where
we can be confident that total monthly precipitation will not exceed
250~mm, which comprises most of the eastern part of the state.

%s5 #&#
\section{Case study 2: Regional climate projections}\label{secApplication-2}

\begin{figure}

\includegraphics{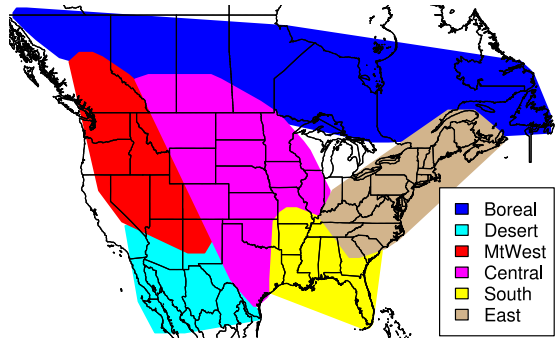}

\caption{Individual NARCCAP regions analyzed.}\label
{figIndividual-NARCCAP-regions}
\end{figure}

We continue to demonstrate the proposed methodology by using it to
explore similarities and differences between projections of
climate models from the North American Regional Climate Change Assessment
Program [NARCCAP; \citet{2011NARCCAPdata}, updated (2012)]. NARCCAP is an international,
multi-disciplinary program exploring ``separate and combined uncertainties
in regional projections of future climate change resulting from the
use of multiple atmosphere-ocean general circulation models (AOGCMs)
to drive multiple regional climate models (RCMs)'' as well as ``to
provide the climate impacts and adaptation community with high-resolution
regional climate change scenarios that can be used for studies of
the societal impacts of climate change and possible adaptation strategies''
[\citet{Mearns2009NARCCAP}; \citet{Mearns2012NARCCAP}]. Data produced by
the program are available
for numerous combinations of AOGCMS and RCMs, allowing researchers
to investigate and study how various models interact, compare, and
contrast with each other. Subsequent analysis will focus on combinations
of two AOGCMs (CCSM and CGCM3) and four RCMs (CRCM, MM5I, RCM3, and
WRFG) with a total of six models considered (CRCM/CCSM, CRCM/CGCM3,
MM5I/CCSM, RCM3/CGCM3, WRFG/CCSM, and WRFG/CGCM3). We consider seasonal
averages (Dec--Feb, Mar--May, Jun--Aug, Sep--Nov) of temperature (degrees
Celsius) for years between 1971 and 2000 (note that the months are
consecutive so that the December of the previous year is included
in the average for the current year) on a 50 km grid covering Canada,
the United States, and the northern part of Mexico. Projections from
each model are also available for years between 2041 and 2070. Potential
predictors to capture large-scale spatial trends include longitude,
latitude, and elevation. The four main goals of this case study are
to:
\begin{longlist}[1.]
\item[1.] Compare and contrast the different AOGCM/RCM models to determine whether
they portray the same type of behavior.
\item[2.] Explore the impact of season on climate predictions.
\item[3.] Study the effect of changing the threshold level on the associated
exceedance region.
\item[4.] Assess where temperature increase is likely to occur based on these
models.
\end{longlist}
The size of the NARCCAP spatial grid and the wide variety of climatological
conditions across the domain necessitated choices for how the data
were analyzed. In order to capture the potential temperature increase
between current and future model runs, future and current model runs
were paired by differences of 70 years (e.g., the average winter temperature
for 1971 was paired with the average winter temperature of 2041) and
the difference between the future and current temperature values were
taken. Depending on how the various RCMs were run, this left between
29 to 30 years of differences for each season. To simplify this part
of the analysis, the data were separated into 9 year groupings
(1971--1979/2041--2049,
1980--1988/2050--2058, and 1989--1997/2059--2067), and the average temperature
difference for each season was calculated. Further, several smaller
and climatologically consistent regions of the domain were analyzed
individually to proceed with analysis. Examples of these smaller regions,
Boreal, Central, Desert, East, MtWest, and South, are overlaid on
a plot of North America in Figure \ref{figIndividual-NARCCAP-regions}.
Using the averages from the three groupings of temperature, statistical
inference for each region proceeded by predicting the average difference
between seasonal temperature between the years 1998--2007 and 2068--2077,
and constructing confidence regions for the exceedance regions of
certain temperature thresholds. For the purposes of this paper, we
will focus on results from the Boreal, South, and East regions
during the winter and summer seasons.

%f6 #&#

Each climate model was analyzed in the following manner. The mean
structure of the responses $Z(\mathbf{s},t)$ was assumed to be constant
across nine-year groupings so that $\mu(\mathbf{s},t)=\mathbf{x}(\mathbf
{s},t)^{\t}\bm{\beta}$,
where the covariates vector $\mathbf{x}(\mathbf{s},t)=[1,\mathrm
{longitude},\mathrm{latitude},\mathrm{elevation}]^{\t}$.
The covariance function of the observed data was modeled as being
isotropic and fully separable with respect to space and time, using
an exponential covariance structure for the spatial covariance and
an $\operatorname{AR}(1)$ structure for the temporal covariance. Measurement error
was assumed to have the same variance for all of the nine-year groupings.
The covariance function of the observed data may be written as
$C(\mathbf{s}_{1}-\mathbf{s}_{2},t_{1}-t_{2})=\sigma^{2}\exp(-h/\phi
)\rho^{u}+\sigma_{\varepsilon}^{2}\mathbf{1}_{(0,0)}(h,u)$,
where $h=\Vert\mathbf{s}_{1}-\mathbf{s}_{2}\Vert$ is the Euclidean distance
between the two data locations and $u=|t_{1}-t_{2}|$ is the time
lag between $t_{1}$ and~$t_{2}$. Restricted maximum likelihood estimation
was used to estimate the covariance parameters $\bm{\theta}=[\sigma
^{2},\phi,\rho,\sigma_{\varepsilon}^{2}]^{\t}$
assuming a multivariate Gaussian distribution for the observed responses.

The next step in our analysis was creating a discrete grid in the
region of interest~$D$, where $D$ is the individual region in question.
Over each domain, a $100\times100$ grid of regular pixels was overlaid,
and then the center points of the pixels within the convex hull bounding
the domain were retained for use as prediction locations. Following
the procedure outlined in Section~\ref{subEstimating-the-Critical-Value},
confidence regions for the exceedance regions of the temperature change
were constructed for levels $u=1$, 2, and 3$^{\circ}$C using 10,000
realizations of the conditional random field in \eqref{eqconditional-sim}.
For each region and exceedance level, both the region $S_{u^{+}}$
containing the true exceedance region $E_{u^{+}}^{Z}$ and the region
$S_{u^{-}}^{c}$ for which all points in the region should be part
of the exceedance region were obtained. The region $S_{u^{+}}$ is
shown in blue and the region $S_{u^{-}}^{c}$ is shown in orange.
Note that because of overlap, all orange locations are covering blue.

\begin{figure}
\centering
\begin{tabular}{@{}c@{}}

\includegraphics{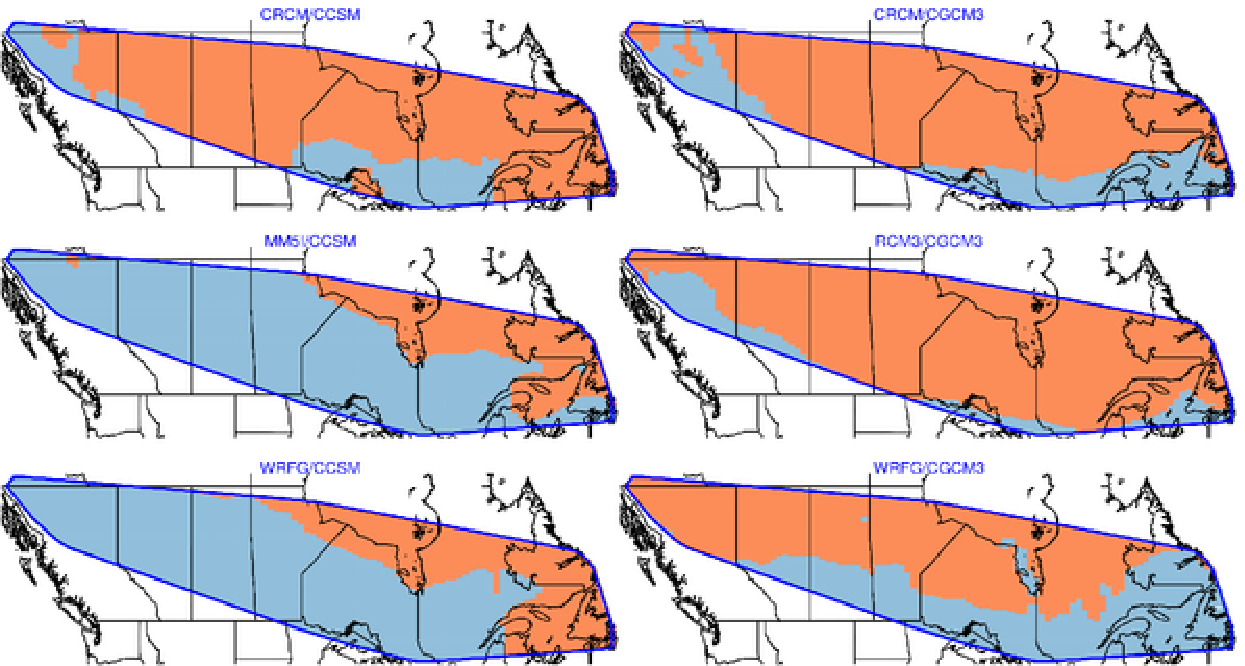}
\\
\footnotesize{(a)}\\[3pt]

\includegraphics{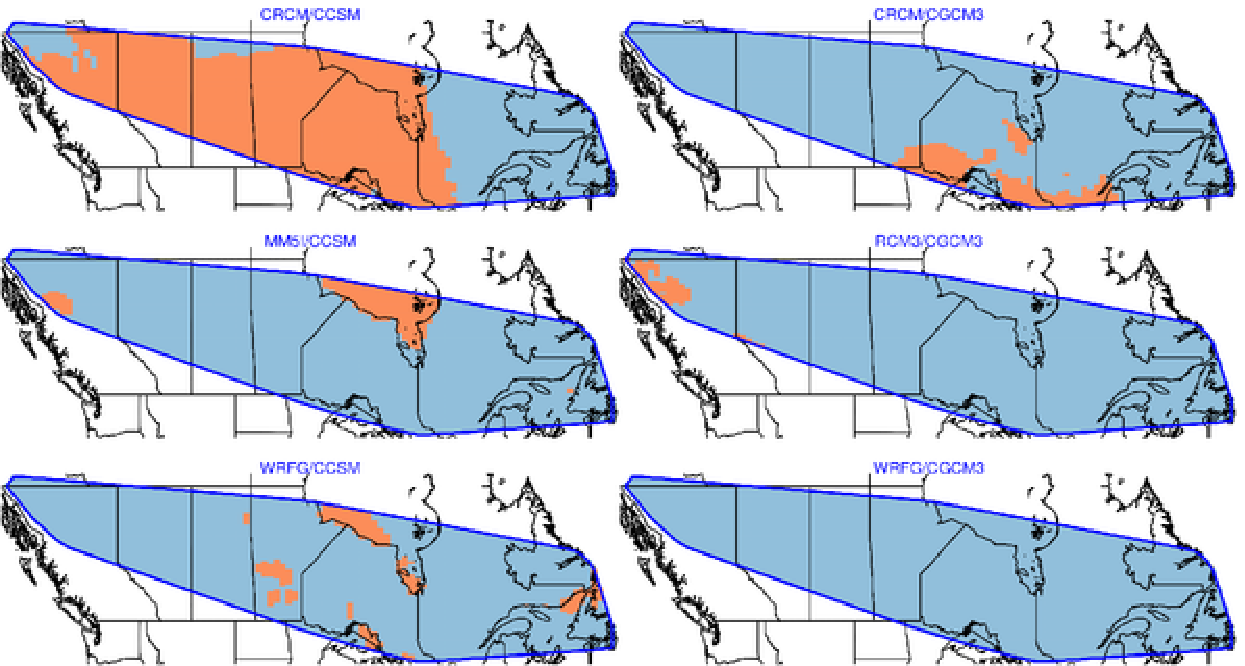}
\\
\footnotesize{(b)}
\end{tabular}
\caption{Confidence regions for locations in the Boreal region where temperature
increase will be at least 1$^{\circ}$C for various climate models.
The results for winter are shown in \textup{(a)} and for summer in \textup{(b)}. Blue
shading indicates where the temperature increase may occur. Orange
shading indicates where the temperature increase is likely to
occur.}\label{figboreal-winter-1}
\end{figure}

We begin by examining the effects of season on temperature predictions
and the associated confidence regions for the Boreal region during
the winter and summer seasons. The confidence regions associated with
temperature change of at least $1^{\circ}$C in winter are shown
in Figure \ref{figboreal-winter-1}(a), and for summer in (b).

%f7 #&#

%}
%temperature
%increase will be at least 1$^{\circ}$ C for various climate models.
%The results for winter are shown in (a) and for summer in (b). Blue
%shading indicates where the temperature increase may occur. Orange
%shading indicates where the temperature increase is likely to occur.

Comparing the graphs in Figure \ref{figboreal-winter-1}(a) and
(b), the pattern of climate change is not necessarily consistent between
winter and summer. Specifically, while the orange area (the area that
we are confident will be part of the true exceedance region) for the
winter data in Figure \ref{figboreal-winter-1}(a) is consistently
in the northeastern area of the Boreal region, we do not see a similar
pattern for the summer data in Figure~\ref{figboreal-winter-1}(b).
The areas that we can confidently identify as experiencing temperature
increase in one season may not experience the same change in a different
season. The confidence region $S_{1^{+}}$ (blue + orange) for the
true exceedance set $E_{1^{+}}^{Z}$ during both winter and summer
comprises the entire Boreal region. Consequently, any part of the
Boreal region could experience a temperature increase of at least
1$^{\circ}$C in both winter and summer. For the winter data, the
orange region consistently makes up a large percentage of the northeastern
Boreal region. Based on the agreement between these climate models,
there appears to be high confidence that the northeastern part of
the Boreal region will experience a temperature increase of at least
1$^{\circ}$C when comparing the average winter temperature between
the years 1989--1997 and 2059--2067. On the other hand, while several
orange regions appear for the summer temperature data in Figure \ref
{figboreal-winter-1}(b), there does not appear to be consistency between the various climate
models. The CRCM/CCSM model results (upper left) in Figure \ref
{figboreal-winter-1}(b) indicate that a majority of the region will experience temperature
change of a least 1$^{\circ}$C during the specified time period,
while the WRFG/CGCM3 model does not indicate that any of the region
will confidently experience this same change. While individual climate
models may deem a certain region as being likely to experience a certain
level of temperature increase, other climate models may paint a different
picture.

\begin{figure}[t!]
\centering
\begin{tabular}{@{}c@{}}

\includegraphics{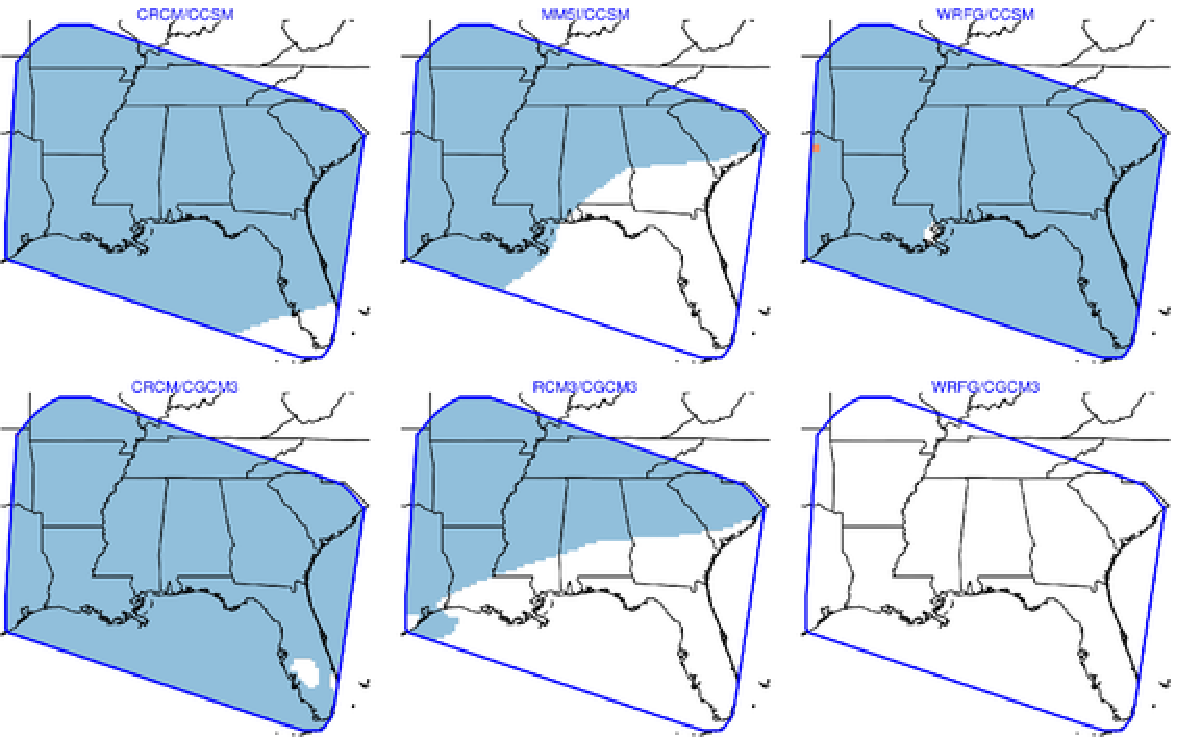}
\\
\footnotesize{(a)}\\[3pt]

\includegraphics{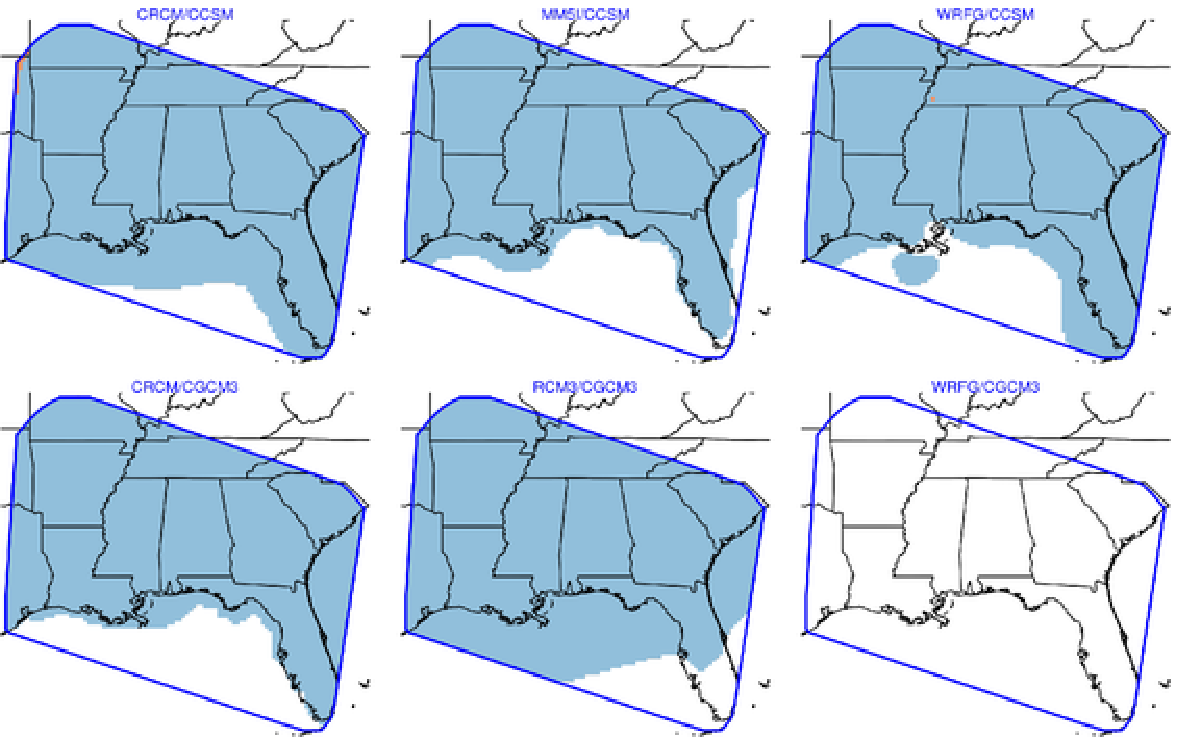}
\\
\footnotesize{(b)}
\end{tabular}
\caption{Confidence regions for locations in the South region where temperature
increase will be at least 3$^{\circ}$C for various climate models.
The results for winter are shown in \textup{(a)} and for summer in \textup{(b)}. Blue
shading indicates where the temperature increase may occur. Areas
without shading are regions where the temperature increase is unlikely
to occur.} \label{figsouth-winter-3}\vspace*{12pt}
\end{figure}

We next consider temperature change in the South region, which comprises
much of the southeastern part of the United States, and focus on how
different RCM/AOGCM climate model combinations can make fairly different
predictions. The results for temperature change of at least 3$^{\circ}$C
during winter and summer are shown in Figure \ref{figsouth-winter-3}(a) and (b), respectively.

%f8 #&#

%
%}
%
%
%}
%
%temperature
%increase will be at least 3$^{\circ}$ C for various climate models.
%The results for winter are shown in (a) and for summer in (b). Blue
%shading indicates where the temperature increase may occur. Areas
%without shading are regions where the temperature increase is unlikely
%to occur. \label{figsouth-winter-3}}

In the South region, the potential temperature increase portrayed
by the climate models is fairly consistent from winter to summer.
Additionally, the results for each of the climate models appear relatively
similar with the exception of the WRFG/CGCM3 combination. Specifically,
this combination predicts temperature increase in the South region
as being smaller than each of the other climate models. This difference
cannot be explained by the use of the CGCM3 AOGCM or WRFG RCM alone,
since other model combinations having these model components do not
behave in the same manner. This specific RCM/AOGCM combination interacts
in a way that yields fairly different results than the other combinations
considered. We note that the nonshaded regions are the regions confidently
identified as not having a temperature increase of $3^{\circ}$C
or more.

\begin{figure}[t!]
\centering
\begin{tabular}{@{}c@{}}

\includegraphics{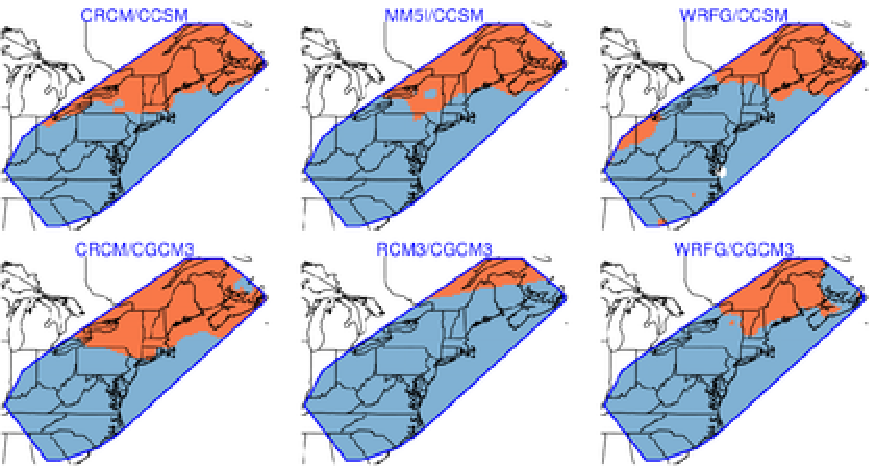}
\\
\footnotesize{(a)}\\[3pt]

\includegraphics{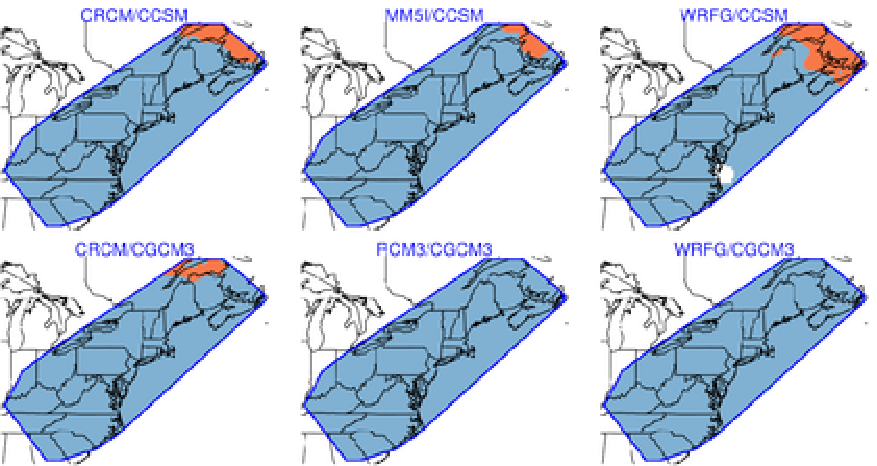}
\\
\footnotesize{(b)}\\[3pt]

\includegraphics{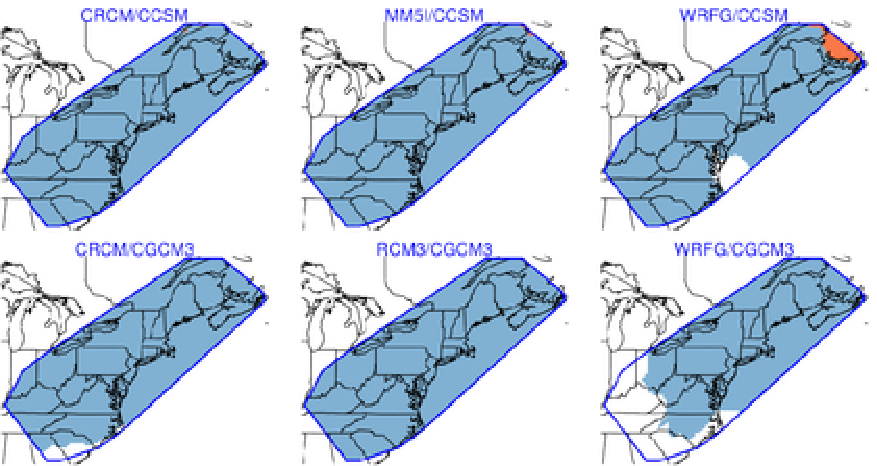}
\\
\footnotesize{(c)}
\end{tabular}
\caption{Confidence regions for locations in the East region where
winter temperature
increase will be at least 1$^{\circ}$C, 2$^{\circ}$C, and 3$^{\circ}$C
are shown in \textup{(a)}, \textup{(b)}, and \textup{(c)}, respectively. Blue shading indicates
where the temperature increase may occur. Orange shading indicates
where the temperature increase is likely to occur. Areas without shading
are regions where the temperature increase is unlikely to occur.}\label
{figeast-winter}
\end{figure}

Last, we consider the effect of changing the threshold level on
the resulting confidence region for the exceedance region by looking
at results for the East region along the Eastern seaboard of the United
States and Canada. Results for winter temperature increases of 1,
2, and 3$^{\circ}$C are shown in Figure \ref{figeast-winter}(a),
(b), and (c), respectively. For all three thresholds, only the northeast
part of the East region confidently has temperature increases exceeding
the threshold in question (the areas shown in orange). However, as
the temperature threshold increases, the size of the orange area decreases.
This behavior is sensible since the more extreme a threshold is, the
less likely it is that a response will exceed that threshold. Consequently,
the orange areas where we are confident a response exceeds a threshold
become smaller as the threshold increases. Based on the observation
that the orange and blue shaded regions make up nearly the entire
East region for all three thresholds, nearly any part of the East
region of North America may experience an average temperature increase
of at least 3$^{\circ}$C when comparing the winter temperature between
the years 1989--1997 and 2059--2067. On the other hand, since the orange
area is mostly limited to the northeastern part of the East region
for a threshold of 1$^{\circ}$C, this is the only region the climate
models can confidently identify as experiencing a temperature increase
of at least 1$^{\circ}$C. On the other hand, as the temperature
threshold is increased to $3^{\circ}$C, some areas have no shading.
These are the regions that can be confidently identified as not experiencing
a temperature increase of $3^{\circ}$C or more. However, the nonshaded
regions are not consistent between the various AOGCM/RCM combinations.

%f9 #&#

%
%}
%
%
%}
%
%
%}
%
%winter temperature
%increase will be at least 1$^{\circ}$ C, 2$^{\circ}$ C, and 3$^{\circ}$
%C are shown in (a), (b), and (c), respectively. Blue shading indicates
%where the temperature increase may occur. Orange shading indicates
%where the temperature increase is likely to occur. Areas without
%shading
%are regions where the temperature increase is unlikely to occur.

We point out that the large blue regions in the preceding discussion
may not be precise indicators of the size of the true exceedance region.
The large size of the blue regions may simply be an indication of
high uncertainty, since the size of the confidence regions $S_{u^{+}}$
will increase with elevated uncertainty.

%s6 #&#
\section{Discussion}\label{secDiscussion}

We have presented an approach for constructing confidence regions
containing the entire exceedance region of a random field. Additionally,
by considering the inverse problem, one may also obtain a region where
all locations in the region will confidently be part of the exceedance
region of the random field. This allows researchers to compare regions
where it is \emph{possible} for an extreme event to occur to the regions
where it is \emph{likely} that an extreme event will occur. The size
of the confidence region naturally depends on factors such as the
number of observed responses, the spatial and temporal dependence,
and the magnitude of measurement error variance. Simulation experiments
in Section \ref{subEmpirical-Confidence-Levels} indicate that the
procedure produces confidence regions having the appropriate coverage
properties. Though all of these experiments used stationary latent
processes for simplicity, stationarity is not required and simulation
experiments for nonstationary processes have produced similar results.
As further explored in Section \ref{subShapes}, the shape of a confidence
region naturally patterns itself after the underlying mean structure.

Using the proposed procedure, we were able to make inference about
the regions of Oregon receiving 250~mm of precipitation in October
of 1998. It also allowed us to compare climate models and explore
future climate change for several combinations of RCMs and AOGCMs
using models from NARCCAP. Though the statistical models used were
basic, the results from these models supported the view that temperature
increases of several degrees are possible for large parts of North
America, and that certain areas seem likely to experience temperature
change of several degrees. The analysis also revealed that there can
be somewhat large discrepancies between climate models (the South
region being a clear example).

It should be mentioned that the assumption that the spatio-temporal
covariance functions of the statistical models were separable may
be unnecessarily restrictive. Nonseparable and/or nonstationary spatio-temporal
covariance models such as the ones proposed by \citet
{Gneiting2002Nonseparable}
or \citet{Fuentes2008Nonseparable} are possible alternatives. Additionally,
the confidence regions were often quite large. This often was a consequence
of the fact that most of the region in question was predicted to be
greater than the exceedance threshold, but was sometimes a result
of high predictive uncertainty in the areas. This suggests that similar
tools with less stringent error criteria [e.g., \citet{Sun2012False}]
would be useful for climate model exploration. Related to this is
the fact that more data brings more information and, consequently,
may reduce the size of the confidence regions. Due to the size and
complexity of the NARCCAP model output, specific regions of North
America were analyzed in isolation. If the proposed procedure were
extended to incorporate reduced rank modeling procedures such as fixed-rank
kriging [\citet{Cressie2008FRK}] or fixed-rank filtering [\citet{Cressie2010FRF}],
this limitation could be ameliorated, and this is the subject of ongoing
research efforts. Additionally, it was pointed out in Section \ref
{subAdditional-Details}
that the confidence levels of the proposed methodology assumed that
the covariance function of the random process was known, which is
rarely the case. A Bayesian approach to this problem that naturally
incorporates the uncertainty of the covariance function is under investigation.

\section*{Acknowledgments}
We wish to thank the North American Regional Climate Change Assessment
Program (NARCCAP) for providing the data used in this paper. NARCCAP
is funded by the National Science Foundation (NSF), the U.S. Department
of Energy (DoE), the National Oceanic and Atmospheric Administration
(NOAA), and the U.S. Environmental Protection Agency Office of Research
and Development (EPA). The National Center for Atmospheric Research is
managed by the University Corporation for
Atmospheric Research under the sponsorship of the National Science
Foundation. The comments of a reviewer and Associate Editor
led to a greatly improved manuscript.

% imsref loaded by akundreckaite, 2013-05-03 14:58:56
%
% imsref loaded by akundreckaite, 2013-05-06 12:31:10


%suskaldyti doi

\printaddresses

\end{document}